\RequirePackage[2020-02-02]{latexrelease}

\documentclass[aps,prl,reprint,superscriptaddress,longbibliography]{revtex4-2}

\usepackage[utf8]{inputenc}
\usepackage{amssymb}
\usepackage{amsmath,bm}
\usepackage{braket}
\usepackage{verbatim}
\usepackage{graphicx}
\usepackage{mathtools}
\usepackage{enumitem}
\usepackage{booktabs}

\usepackage[dvipsnames]{xcolor}

\usepackage{hyperref}
\hypersetup{%
	pdftoolbar=true,        
	pdfmenubar=true,        
	pdffitwindow=false,     
	pdfstartview={FitH},    
	pdftitle={title},    
	pdfauthor={Gerard Valentí-Rojas},     
	pdfsubject={Subject},   
	pdfcreator={Gerard Valentí-Rojas},   
	pdfproducer={Gerard Valentí-Rojas}, 
	pdfnewwindow=true,      
	colorlinks=true,       
	linkcolor=RoyalBlue,          
	citecolor=OrangeRed,        
	filecolor=Green,      
	urlcolor=RoyalBlue           
}

\usepackage{newtxtext,newtxmath}

\begin{document}


\title{Topological Gauge Fields and the Composite Particle Duality
}


\author{Gerard Valentí-Rojas}
\email[Corresponding author: ]{gv16@hw.ac.uk}
\affiliation{SUPA, Institute of Photonics and Quantum Sciences, Heriot-Watt University, Edinburgh, EH14 4AS, United Kingdom}

\author{Aneirin J. Baker}
\affiliation{SUPA, Institute of Photonics and Quantum Sciences, Heriot-Watt University, Edinburgh, EH14 4AS, United Kingdom}

\author{Alessio Celi}
\affiliation{Departament de Física, Universitat Autònoma de Barcelona, 08193 Bellaterra, Spain}

\author{Patrik Öhberg}
\affiliation{SUPA, Institute of Photonics and Quantum Sciences, Heriot-Watt University, Edinburgh, EH14 4AS, United Kingdom}


\date{\today}

\begin{abstract}
We introduce topological gauge fields as nontrivial field configurations enforced by topological currents. These fields crucially determine the form of statistical gauge fields that couple to matter and transmute their statistics. We discuss the physical mechanism underlying the composite particle picture and argue that it is a duality of gauge forms that naturally relates to the notion of bosonisation in arbitrary dimensions. This is based on obtaining a generalised version of flux attachment, which yields a density-dependent gauge potential. We recover well-known results, resolve old controversies, and suggest a microscopic mechanism for the emergence of such a gauge field. We also outline potential directions for experimental realisations in ultracold atom platforms.

\end{abstract}


\maketitle


\paragraph*{\textit{Introduction.}}


Flux attachment \cite{wilckez1982flux} is a physical mechanism describing how charged particles capture magnetic flux quanta and become composite entities, often featuring exotic properties \cite{wilczek1982fraction}. In planar (2+1D) systems, this constitutes a well-established picture to intuitively understand the low-energy effective description of some topologically ordered phases of matter \cite{wen1990topological,moessner2021topological}. The appearance of a Chern-Simons field is found responsible for this phenomenon and is intrinsically related to the fractionalisation of quantum numbers \cite{feldman2021fractional,forte1992review}. All the above can be encapsulated as part of a Bose-Fermi correspondence in which dynamical gauge fields play a pivotal role \cite{polyakov1988fermi}.
The situation is much different in other dimensions, where the previous concepts become ill-defined. For instance, there is no Chern-Simons term in even spacetime dimensions, so the existence of point-like anyons \cite{leinaas1977theory}, as well as their interpretation as composites, seems no longer valid. It is then natural to ask whether an analogous mechanism to flux attachment can be found in all generality. This is a subtle question, especially in linear (1+1D) systems, where the notion of a magnetic field is not even defined so, strictly speaking, there is no flux to attach. Yet, Bose-Fermi correspondences \cite{jordan1928paulische,coleman1975sine,mandelstam1975soliton,valiente2021universal} seem almost inevitable and many instances of linear anyons have been proposed \cite{Harshman2022exchange}.

In this work, we introduce a \textit{composite particle duality} understood as an extended notion of flux attachment in arbitrary dimensions (see Table \ref{table:sum} for a conceptual summary). The correspondence should be thought of as physically gauge dressing a set of charged fields. Matter becomes electromagnetically charged by construction and, in a gauge transformed or dual picture, is seen as composites. The power of our result relies on its universality as it holds regardless of the initial system being originally bosonic or fermionic, continuous or in the lattice, as long as matter interactions remain short-ranged. As an application, we then study an illustrative non-relativistic model in one spatial dimension. We verify the presence of fractional statistics and resolve an old controversy \cite{rabello1996prl,aglietti1996solitons,kundy99anyons}. In this context, and based on recent experimental work in ultracold atoms \cite{frolian2022realising,chisholm2022encoding}, we suggest a microscopic origin for such a statistical gauge field emerging in interacting quantum many-body systems.
More broadly, evidence from the aforementioned duality enables us to make  the following general observations: 

\textit{(i)} The notion of a composite particle duality naturally generalises those of conventional flux attachment and statistical transmutation. It survives in any dimension, its origin is geometric, it is physically enforced by topological terms, and satisfies an \textit{order-disorder} operator structure \cite{marino2017quantum}. 

\textit{(ii)} Such Bose-Fermi correspondences can be probed in experiments by tuning the coupling constant of the statistical gauge field or detecting gauge-charge composite quasiparticles. Topological terms in the effective action can arise from conventional interparticle interactions, showing that the associated statistical gauge fields can be physical as opposed to fictitious or merely formal, as often regarded in literature \cite{wilczek1990book}.

Finally, we speculate that D-dimensional Abelian \textit{bosonisation} might be understood as \textit{statistical transmutation}, a consequence of the coupling of matter to statistical gauge fields. This would be crucially related to Jordan-Wigner-like mappings interpreted either as large gauge transformations or as attaching topological gauge defects in the form of disorder operators. 

\begin{table}[ht]
	\centering
	\setlength{\tabcolsep}{2pt}
	\renewcommand\arraystretch{2}
	\begin{tabular}[t]{l  c  c  c}
		\toprule
		\textit{Spatial Dimension}&$d=1$&$d=2$&$d=3$\\
		\midrule
		\textit{Flux Attachment}&$a_{x} \propto n$&$b \propto n$&$\bm{\nabla}\cdot \bm{b} \propto n$\\
		\textit{``Hall" Response}&$a_{t}\propto J_{x}$&$\hat{\mathbf{e}}_{z}\times \bm{\varepsilon} \propto \bm{J}$&$\partial_{t}\bm{b} + \bm{\nabla}\times \bm{\varepsilon} \propto \bm{J}$\\
		\textit{Top. Quantisation}&No&Yes&Yes\\\vspace*{-2.1mm}
		\textit{Top. Soliton}&Kink&Vortex&Monopole\\ \vspace*{-3.2mm}
		\textit{Statistical Field}&\parbox[c]{5em}{\includegraphics[scale=0.028]{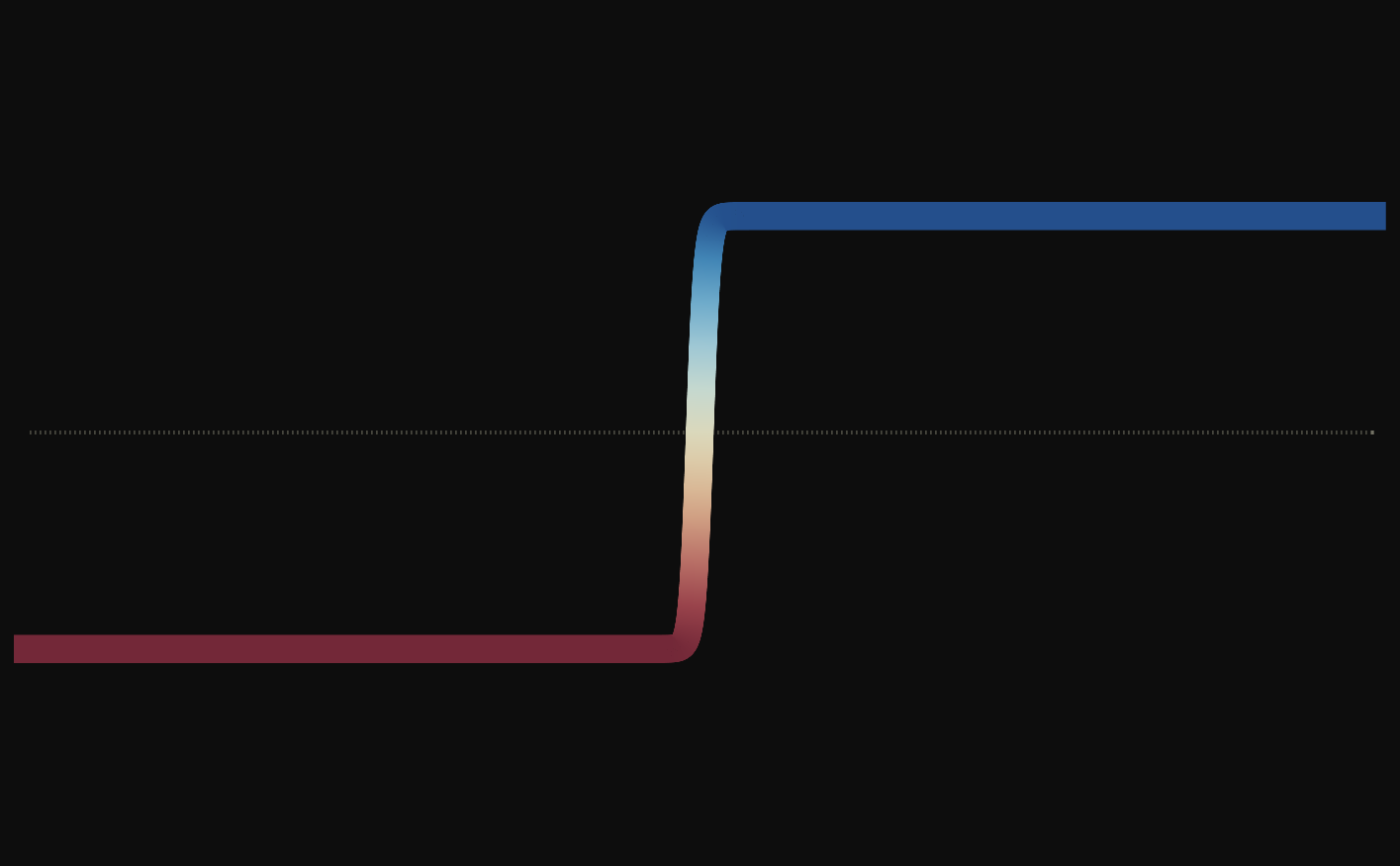} }
		&\parbox[c]{5.8em}{\includegraphics[scale=0.044]{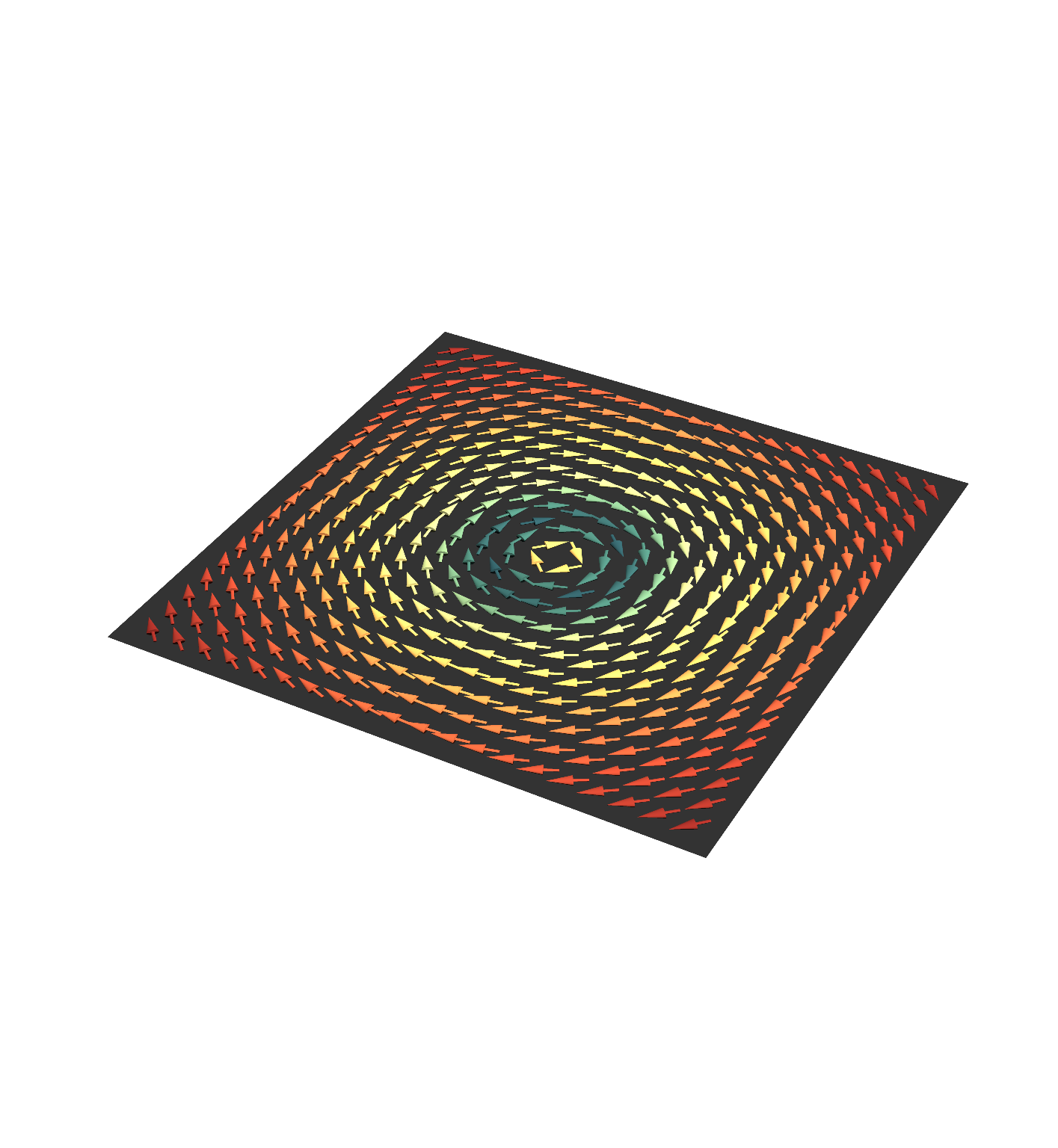} }&\parbox[c]{6.5em}{\includegraphics[scale=0.04]{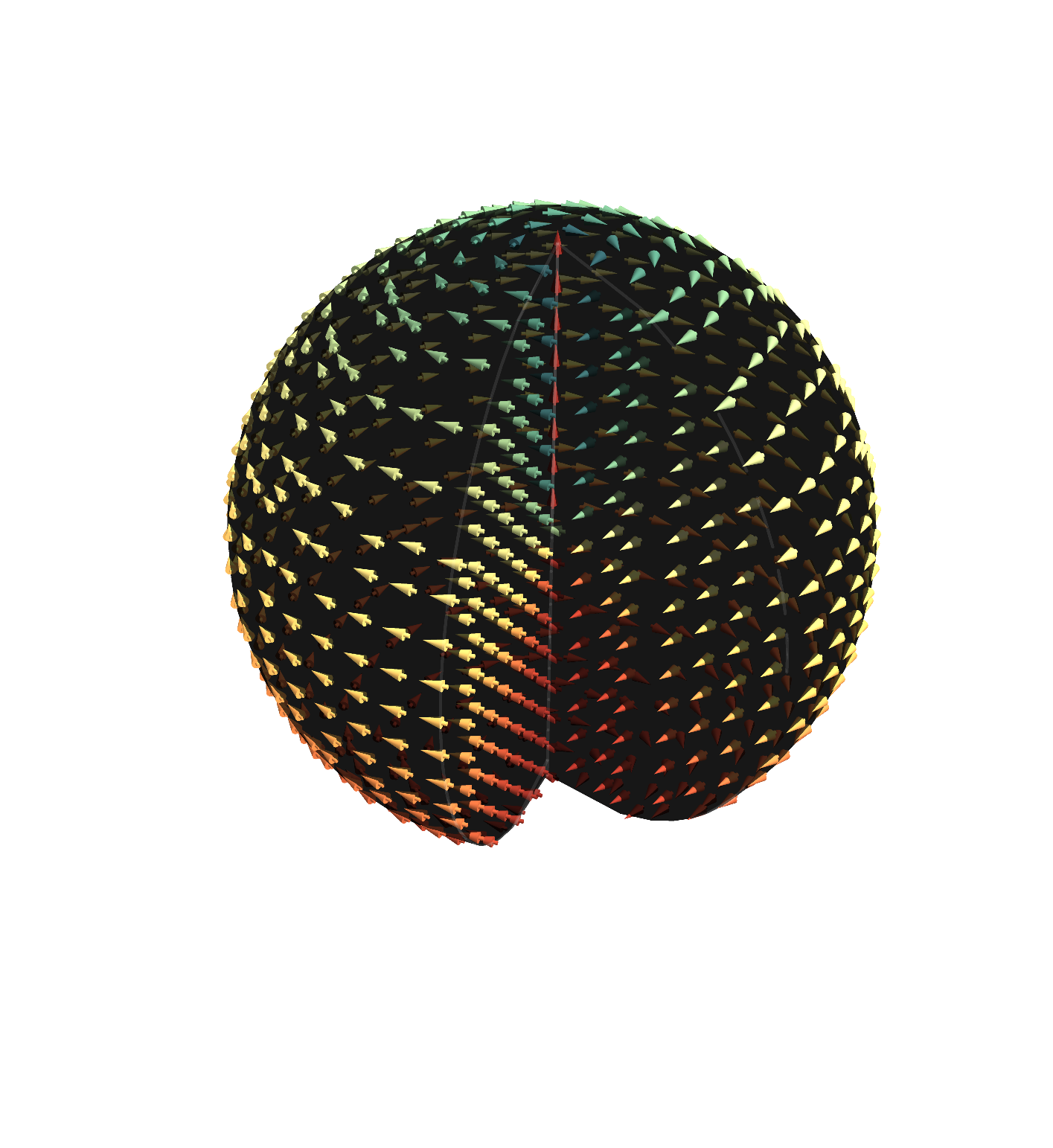} }\\ 
		\bottomrule
	\end{tabular}
\caption{Main features of the electromagnetic response for matter coupled to statistical gauge fields. Blue (red) colouring at the bottom row denotes high (low) intensity of the vector field. Observe a jump discontinuity at the origin for the kink, a singular point for the vortex and a singular Dirac string on the positive $z-$axis for the monopole.}
\label{table:sum}	
\end{table}%

\paragraph*{\textit{Composite Particle Duality.}}
We establish that the minimal coupling of charged matter to a \textit{statistical} gauge field $a_{\mu}$ in $\text{D}=d+1$ spacetime dimensions is equivalent to the formation of electric-magnetic entities identified as gauge-charge composites. In some instances, the latter may be regarded as anyons. 
This correspondence is summarised as
\begin{equation}\label{eq:comp_part_duality}
\mathcal{H}_{\text{\,B}} = \sum_{i=1}^{N} \frac{\bm{\pi}_{i}^{\,2}}{2m} + \mathcal{H}_{\text{int}}\;\; \longleftrightarrow \;\; \tilde{\mathcal{H}}_{\text{\,C}} = \sum_{i=1}^{N}\frac{\tilde{\bm{p}}_{i}^{\,2}}{2m} + \tilde{\mathcal{H}}_{\text{int}}\;\;,
\end{equation}
where $\bm{\pi}_{i} = \bm{p}_{i} - \bm{a} (\mathbf{x}_{i})\,$, $N$ is the number of particles, and interactions are short-ranged. We postulate that the statistical gauge potential is a topologically non-trivial pure gauge configuration
\begin{equation}\label{eq:pure_gauge}
	\bm{a}(\mathbf{x}_{i})  = \bm{\nabla}_{\mathbf{x}_{i}} \Phi\,(\mathbf{x}_{1},\dots,\mathbf{x}_{N})\;,
\end{equation}	
where $\bm{a}(\mathbf{x}_{i}) \equiv \bm{a}(\mathbf{x}_{i}\,;\,\mathbf{x}_{1},\dots,\mathbf{x}_{N}) $ refers to the gauge potential being evaluated at the location of particle $\mathbf{x}_{i}$, although it might be a function of the position of all the particles in the system and, as we will find later, also of matter density $\lvert \Psi_{\text{B}}\,(\mathbf{x}_{1},\dots,\mathbf{x}_{N})\rvert^{2}\,$. We identify a many-body Hamiltonian in the \textit{bare} (B) basis, and another corresponding to the \textit{composite} (C) basis. Both sides of the duality are related by a large gauge transformation which removes/introduces a minimally coupled statistical gauge field. This transformation corresponds to the naive generalised continuum version of the well-known Jordan-Wigner transformation which, in second-quantised language, reads
\begin{equation}\label{eq:jw_transf}
\hat{\Psi}_{\text{C}}\,(\mathbf{x}; \Gamma_{\mathbf{x}}) =	\hat{\mathcal{W}}^{\,\dagger} (\mathbf{x}; \Gamma_{\mathbf{x}}) \,\hat{\Psi}_{\text{B}}\,(\mathbf{x}) \;.
\end{equation}
The operator $\hat{\mathcal{W}}\,(\mathbf{x};\Gamma_{\mathbf{x}})= \exp \,\big[i\hbar^{-1}\hat{\Phi}\,(\mathbf{x}; \Gamma_{\mathbf{x}})\big]$ is identified as a\textit{ disorder} operator \cite{kadanoffceva1971,fradkin2017disorder}, $\hat{\Phi}$ is a $(\text{D}-1)-$dimensional \textit{Jordan-Wigner brane} \footnote{This is meant to generalise the concept of a Jordan-Wigner string to higher dimensions. For low dimensions the \textit{brane}, and thus the composite operator, are local, i.e. they can be defined at a given point in space. However, for $\text{D}\ge 3$ this is not the case and these objects become intrinsically non-local, a feature captured by $\Gamma_{\mathbf{x}}$.}, and $\Gamma_{\mathbf{x}}$ is a reference contour centred at $\mathbf{x}$ in the sense of \cite{marino2017quantum}. Considering the bare species $\hat{\Psi}_{\text{B}}$ to be a bosonic (fermionic) field, and hence satisfying ordinary equal-time (anti)commutation  relations, then, $\hat{\Psi}_{\text{C}}$ constitutes a composite field obeying generalised commutation relations, except for at the point $\mathbf{x} = \mathbf{x'}$, where relations reduce to those of bare species due to cancellation of branes. The density operator is $\hat{n} (\mathbf{x}) = \hat{\Psi}_{\text{B}}^{\dagger} (\mathbf{x})\,\hat{\Psi}_{\text{B}}(\mathbf{x}) = \hat{\Psi}_{\text{C}}^{\dagger}(\mathbf{x};\Gamma_{\mathbf{x}}) \,\hat{\Psi}_{\text{C}}(\mathbf{x};\Gamma_{\mathbf{x}})\,$, so all possible local interaction terms which are functions of the density are identical on both sides of the duality. 

The explicit form of the statistical gauge field might be guessed, but it can also be derived on the grounds of topological field theory. Hence, we introduce a \textit{topological} gauge field $\hat{\mathcal{A}}$ as a new gauge connection obtained from solving a topological current constraint satisfying $\partial_{\mu} \hat{\mathcal{J}}^{\mu} (\mathbf{x}) = 0$ by construction. A particularly simple form is 
\begin{flalign}\label{eq:topo_current}
	&\hat{\mathcal{J}}^{\,\mu} = \frac{\kappa}{2\pi} \,\epsilon^{\,\mu \nu \lambda \alpha \beta \dots } \,\partial_{\nu}\, \hat{\mathcal{A}}_{\,\lambda \alpha \beta \dots}\;\;,
\end{flalign}
where $\hat{\mathcal{A}}\,$ is an antisymmetric gauge-dependent $(\text{D}-2)$--form field that transforms like $\hat{\mathcal{A}}_{\mu\nu\lambda\alpha\dots} \rightarrow \hat{\mathcal{A}}_{\mu\nu\lambda\alpha\dots} + \partial_{\,[\mu}\,\hat{\xi}_{\,\nu\lambda\alpha\dots]}\,$, and $\kappa$ is a real constant whose value may or may not be constrained through Dirac quantisation. This general relation reduces to a scalar field ($\hat{\Phi}$) in 1+1D, a Chern-Simons field ($\hat{A}_{\mu}$) in 2+1D, and a Kalb-Ramond field ($\hat{B}_{\mu \nu}$) in 3+1D or higher ($\hat{C}_{\mu\nu\lambda\dots}$). A relation such as that in Eq. \eqref{eq:topo_current} has been identified and discussed extensively in previous work on \textit{functional bosonisation} \cite{le1996current,fradkin1994fermion,frohlich1995bosonize,burgess1994bosonization,burgess1994bosonization2,schaposnik1995comment,fosco2018functional,schaposnik1998bosonization,banerjee1997501,chanmain13functional,cirio2014tight} and is nothing but a manifestation of the Hodge duality $\hat{\mathcal{F}} \equiv \text{d}\hat{\mathcal{A}}\,= \star \hat{\mathcal{J}} $ mapping $p-$forms in $\text{D}$ dimensions to $(\text{D}-p)-$forms through the Hodge star operator $(\star)\,$. We can write the $(\text{D}-2)-$form field as $\mathcal{\hat{A}} = \text{d} \hat{\alpha}$ at the expense of violating the corresponding Bianchi identity $\text{d}(\text{d}\hat{\alpha}) \neq 0$ due to non-trivial topology of $\hat{\alpha}\,$, hence the name for $\mathcal{\hat{A}}$. The current constraint can also be found as an equation of motion of a parent effective gauge action when minimally coupled to a matter source. In the following, we restrict ourselves to spatial dimensions $d=1,2,3$. An educated guess for an action is then $S\,[a, \mathcal{A}] = S_{\text{topo.}} - \int \text{d}^{\text{D}}\mathbf{x} \,\mathcal{J}^{\mu}a_{\mu}\,$, where
\begin{equation}\label{eq:topo_action}
S_{\text{topo.}} = \int \text{d}^{\text{D}} \mathbf{x}\;\bigg(\frac{\kappa}{2\pi}\,\epsilon^{\,\mu \nu \lambda \dots \alpha \beta } \mathcal{A}_{\mu \nu \lambda \dots} \partial_{\alpha} a_{\beta} + \mathcal{L}\,[\mathcal{A}] \,\bigg)\;,
\end{equation}
is a combination of Background Field (BF) and another topological term changing with spatial dimension. Upon elimination of the topological gauge field, one recovers Chern-Simons terms for odd D-dimensions or $\theta$--terms for even. These terms will, in general, give restricted dynamics to the statistical gauge field when coupled to matter via a Gauss's law constraint. The statistical gauge field ($a$) is \textit{physically} coupled to matter while the topological gauge field ($\mathcal{A}$) is \textit{auxiliary}. The Euler-Lagrange equations with respect to both define, respectively, the current bosonisation rule and a local condition expressing the topological gauge field in terms of the statistical one.
While we have no apparent evidence to expect the action \eqref{eq:topo_action} to  be unique, we claim that the terms involved should induce torsion or helicity \cite{jackiw2004perfect} (e.g. a twist) in the gauge connection and not merely curvature. 
Upon using the equations of motion, we recover the relations
\begin{flalign}\label{eq:topo_cons_1d}
&\mathcal{J}^{\,\mu}_{\,\text{1+1D}} = \frac{\kappa}{2 \pi}\epsilon^{\,\mu \nu} \partial_{\nu} \Phi = \frac{\kappa}{2 \pi}\epsilon^{\,\mu \nu} a_{\nu}\, ,\\
&\mathcal{J}^{\,\mu}_{\,\text{2+1D}} = \frac{\kappa}{2 \pi}\epsilon^{\,\mu \nu \lambda} \partial_{\nu} A_{\lambda} =  \frac{\kappa}{2 \pi} \epsilon^{\,\mu \nu \lambda} \partial_{\nu} a_{\lambda} \,,\\
&\mathcal{J}^{\,\mu}_{\,\text{3+1D}} =  \frac{\kappa}{2 \pi}\epsilon^{\,\mu \nu \lambda \alpha} \partial_{\nu} B_{\lambda \alpha} =  \frac{\kappa}{2 \pi}\epsilon^{\,\mu \nu \lambda \alpha} \partial_{\nu} \partial_{\lambda} a_{\alpha}\,.
\end{flalign}
From here, we can read the corresponding flux attachment Gauss's laws in vector notation as
\begin{flalign}\label{eq:kink1d}
&(\text{1+1D}) :\;\;\;\;\;\;\;\;\;\;\;\; \frac{2 \pi}{\kappa}\, n\,(t,x) = a_{x} \,(t,x) \\\label{eq:vortex2d}
&(\text{2+1D}) :\;\;\;\;\;\;\;\;\;\;\;\; \frac{2 \pi}{\kappa}\, n\,(t,\bm{x}) = b \,(t,\bm{x})\\ \label{eq:monopole}
&(\text{3+1D}) :\;\;\;\;\;\;\;\;\;\;\;\; \frac{2 \pi}{\kappa}\,n \,(t,\bm{x})= \bm{\nabla} \cdot \bm{b}  \,(t,\bm{x})
\end{flalign}
where the ``electric" charge density is $\mathcal{J}^{\,0} = n\,(t,\bm{x})$ and ``magnetic" field is $\bm{b} \,(t,\bm{x}) \equiv b^{\,i}  = \epsilon^{\,ijk}\partial_{j} a_{k} \,$ \footnote{We use the shorthand notation $\mathbf{x} =(t,\bm{x})$ unless explicitly noted otherwise and the convention that charge or number density is given by $\hat{\mathcal{J}}^{\,0} = \hat{n}$. In doing so we are taking the electric unit of charge $q=c=1$ to be equal to one.}. The 2+1D constraint corresponds to the usual description of flux attachment in the low-energy effective formulation of the fractional quantum Hall effect (FQHE). The 3+1D case is exceptionally striking since Eq. \eqref{eq:monopole} can be immediately identified as the magnetic monopole law, but with $n\,(t,\bm{x})$ being an electric charge density, not the usual magnetic source. In other words, an entity carrying electric charge also becomes a source magnetic field, i.e. it forms a composite reminiscent of a \textit{dyon} \cite{schwinger1969magnetic,schwinger1968sources,schwinger1966charge}. This is indeed the consequence of flux attachment in three spatial dimensions and thus, such entities would appear as three-dimensional analogues of Laughlin quasiparticles.

\paragraph*{\textit{A Simple Dual Model.}}
The correspondence in Eq. \eqref{eq:comp_part_duality} supplemented by the statistical gauge fields found in Eqs. (\ref{eq:kink1d}--\ref{eq:monopole}) and understood as a physical mechanism for statistical transmutation constitutes the main result of this work. Hereafter, for illustrative purposes, we will be interested in the 1+1D case, for which the time component of the topological current in Eq. \eqref{eq:topo_current} reduces to the usual bosonisation relation $\hat{n}\, (\text{x}) = \gamma^{-1} \,\partial_{x}\,\hat{\Phi} \,(\text{x})$ where $\gamma = 2\pi / \kappa\,$. By simple integration, we see that $\hat{\Phi}\, (\text{x}) = \gamma\,\int_{-\infty}^{x} \text{d}x'\;\hat{n}\,(\text{x}')$. This appears naturally for a density-dependent gauge potential of the form
\begin{equation} \label{eq:dens-dep_gauge}
	\hat{a}_{x}\,(\text{x})  = \gamma \int_{-\infty}^{\infty} \text{d}x'\;\partial_{x} \,\Theta\,(x-x') \,\hat{n}\,(\text{x}') = \gamma\,\hat{n}\,(\text{x})\;,
\end{equation}
where $\Theta\,(x)$ denotes a Heaviside step function, which plays the role of a \textit{kink}. We observe that the gauge potential in 1+1D is a pure gauge one, $\hat{a}_{x} = \partial_{x}\hat{\Phi}\,$. 
Provided there is no singularity to wind around, we expect no topological quantisation in this model as opposed to its higher dimensional relatives. Additionally, we notice that for a system of point particles, in first-quantised language, Eq. \eqref{eq:dens-dep_gauge} non-trivially reduces to $a_{x} (\text{x}_{i}) = \partial_{x_{i}} \Phi= \gamma\,\sum_{j\neq i} \text{sgn}\,(j-i)\, \delta\,(x_{i} - x_{j})$ with a string $\Phi = \gamma \sum_{j<l} \Theta\,(x_{j} -x_{l})$. This resolves the long-standing tension between Refs. \cite{rabello1996prl,aglietti1996solitons,kundy99anyons} as it systematically corrects the sign error made in Ref. \cite{rabello1996prl} and identified in Ref. \cite{aglietti1996solitons}. 

In order to understand the implications of the density-dependent gauge field \eqref{eq:dens-dep_gauge} and the duality, we should study its behaviour in the presence of dynamical matter. The discussion that follows can be taken as an extension of Kundu's results \cite{kundy99anyons}.
We consider a 1+1D non-relativistic weakly-interacting Bose gas \footnote{We emphasise that the bare species chosen could also be fermionic and the mechanism would work analogously. Similarly, for a relativistic model such as a complex Klein-Gordon field, our claims still hold. Here we choose a bosonic field for convenience and experimental relevance in ultracold atoms.} minimally coupled to a statistical gauge field with action $S\,[\hat{\Psi},\hat{a}_{\mu}] = \int \text{d}t\,\text{d}x\,\mathcal{L}_{\text{B}}$ and Lagrangian density
\begin{flalign}
 \label{eq:bare}
	&\mathcal{L}_{\text{B}}= i\hbar \,\hat{\Psi}^{\dagger}D_{t}\hat{\Psi} - \frac{\hbar^{2}}{2m} \big(D_{x}\hat{\Psi}\big)^{\dagger} \big(D_{x}\hat{\Psi}\big) - \frac{g}{2}\, \hat{\Psi}^{\dagger}\hat{\Psi}^{\dagger}\hat{\Psi}\hat{\Psi} .
\end{flalign}
The form of the gauge field is given by the topological constraint $\hat{J}^{\,\mu} = \gamma^{-1}\epsilon^{\,\mu\nu}\hat{a}_{\nu}$ in Eq. \eqref{eq:topo_cons_1d}, where $\hat{J}^{\,\mu} = \big(\hat{n}\,, \hat{J}_{x}\big)\,$, which reads
\begin{equation}
\hat{J}^{\,\mu}\,(\text{x}) = \bigg(\,\hat{\Psi}^{\dagger} \hat{\Psi}\,, \;\frac{\hbar}{2mi}\, \Big[\hat{\Psi}^{\dagger}D_{x}\,\hat{\Psi} - (D_{x}\,\hat{\Psi})^{\dagger}\,\hat{\Psi}\Big] \,\bigg)\;,
\end{equation}
is a conserved current. The gauge covariant derivative is given by $D_{\mu} = \partial_{\mu} - i\hbar^{-1} \hat{a}_{\mu}\,$ and the commutation relations are
\begin{flalign}
	&\big[\hat{\Psi}(x)\,,\hat{\Psi}(x') \big] =\big[\hat{\Psi}^{\dagger}(x)\,,\hat{\Psi}^{\dagger}(x') \big]= 0 \;,\\
	&\big[\hat{\Psi}\,(x)\, ,\hat{\Psi}^{\dagger}(x') \big]= \delta\,(x-x')\;.
\end{flalign}
This model admits a composite dual description of the form $ \tilde{S}\,[\hat{\Psi}_{\text{C}}] = \int \text{d}t\,\text{d}x\;\tilde{\mathcal{L}}_{\text{C}}$ with
\begin{flalign}
	&\tilde{\mathcal{L}}_{\text{C}} = i\hbar\, \hat{\Psi}_{\text{C}}^{\dagger}\,\partial_{t} \hat{\Psi}_{\text{C}}  - \frac{\hbar^{2}}{2m} \, \partial_{x} \hat{\Psi}_{\text{C}}^{\dagger} \,\partial_{x} \hat{\Psi}_{\text{C}} -\frac{g}{2} \,\hat{\Psi}_{\text{C}}^{\dagger}\hat{\Psi}_{\text{C}}^{\dagger} \hat{\Psi}_{\text{C}} \hat{\Psi}_{\text{C}}
\end{flalign}
where the composite field obeys the algebra
\begin{flalign}
	&\hat{\Psi}_{\text{C}}^{\dagger}(x) \,\hat{\Psi}_{\text{C}}^{\dagger}(x') - e^{\,\frac{i}{\hbar}\gamma \,\text{sgn}\,(x-x')} \,\hat{\Psi}_{\text{C}}^{\dagger}(x')\, \hat{\Psi}_{\text{C}}^{\dagger}(x) = 0 \\
	&\hat{\Psi}_{\text{C}} (x) \,\hat{\Psi}_{\text{C}}^{\dagger}(x') - e^{\,-\frac{i}{\hbar}\gamma \,\text{sgn}\,(x-x')} \,\hat{\Psi}_{\text{C}}^{\dagger}(x')\, \hat{\Psi}_{\text{C}} (x) = \delta\,(x-x')
\end{flalign}
and the coupling $\gamma$ constitutes also the statistical angle. The composite dual action is found via the Jordan-Wigner transformation [Eq. \eqref{eq:jw_transf}], which reads as
\begin{equation}
\hat{\Psi}\,(\text{x}) = e^{\,\frac{i}{\hbar} \gamma \int_{-\infty}^{\infty} \text{d}x'\;\Theta\,(x-x') \,\hat{n}\,(\text{x}')}\,\hat{\Psi}_{\text{C}}\,(\text{x})
\end{equation}
and acts like a \textit{large} gauge transformation and allows interpolation between both faces of the duality. This constitutes a \textit{statistical transmutation}. Notice that the Jordan-Wigner string is nothing but the aforementioned scalar (kink) field $\hat{\Phi} \,(\text{x})\,$. A crucial rewriting of the kinetic term as
\begin{flalign}
	\tilde{H}_{\text{kin}} &\sim\partial_{x}\hat{\Psi}_{\text{C}}^{\dagger} \,\partial_{x} \hat{\Psi}_{\text{C}} =\big(D_{x} \,\hat{\Psi}\big)^{\dagger} \big(D_{x}\, \hat{\Psi}\big) \\ \label{eq:parity}
	&= \partial_{x} \hat{\Psi}^{\dagger} \partial_{x} \hat{\Psi} -\frac{2\gamma m}{\hbar^{2}} \colon \hat{n}\, \hat{j}_{x} \colon + \frac{\gamma^{2}}{\hbar^{2}} \,\hat{\Psi}^{\dagger} \hat{n}^{2} \,\hat{\Psi} \;,
\end{flalign}	
with $\colon\bullet\,\colon$denoting normal ordering and current density 
\begin{equation}
	\hat{j}_{x} = \frac{\hbar}{2mi}\, \Big[\hat{\Psi}^{\dagger} \partial_{x}\,\hat{\Psi} - \big(\partial_{x}\, \hat{\Psi}\big)^{\dagger} \,\hat{\Psi}\Big]\;,
\end{equation}
shows that minimal coupling to a density-dependent gauge potential is nothing but a density-current nonlinearity and a three-body term. Thus, modifying the coupling of the interactions in Eq. \eqref{eq:parity} is equivalent to tuning the statistics in the dual description. The most dramatic effects are produced by the term involving the current density, which breaks parity ($\mathcal{P}$) and time-reversal ($\mathcal{T}$), and leads to asymmetric dynamics. The latter has been studied at the mean-field level in Ref. \cite{edmonds2013simulating}, where it gives rise to unusual propagation dynamics and the formation of chiral solitons. These features have  now been experimentally confirmed \cite{frolian2022realising,chisholm2022encoding} and studied in more detail. The composite dual model is also known as the anyonic Lieb-Liniger model, see e.g. \cite{calabrese2007correlation, piroli2020determinant, scopa2020one} and references therein. It is worth noting that Girardeau's Bose-Fermi correspondence \cite{girardeau2006anyon,girardeau1960relationship,cheon1999fbduality} is a particular case of the above prescription.

\paragraph*{\textit{Macroscopic Origin.}}
The reader might wonder whether the statistical gauge field in Eq. \eqref{eq:dens-dep_gauge} has really anything to do with conventional flux attachment. Let us consider the local law $\bm{\nabla}\times\hat{\bm{a}}(\bm{x}) = \gamma\, \hat{n} (\bm{x})$ on a punctured 2d disk with  $r=\epsilon$ inner and $r=R$ outer radius, respectively. When taking the limit $\epsilon \rightarrow R$ the disk approaches an annulus. This implies $\lvert R-\epsilon\rvert \approx 0$ and $\partial_{r} \,\hat{\bm{a}}\approx  0\,$ in $\epsilon \le r \le R$. The magnetic field in polar coordinates $(r,\varphi)$ becomes
\begin{equation}
\bm{\nabla}\times\hat{\bm{a}}(\bm{x})\Big\rvert_{\epsilon \rightarrow R} = \Big[ \frac{1}{r}\,\hat{a}_{\varphi}(r,\varphi) - \frac{1}{r}\,\partial_{\varphi} \,\hat{a}_{r}(r,\varphi)\Big] \Big\rvert_{\,r \rightarrow R} \;.
\end{equation}
In this limit, the flux attachment expression at $r=R$ reads
\begin{equation}
\hat{a}'_{\varphi}(\varphi) = \frac{1}{R}\, \Big[\hat{a}_{\varphi}(\varphi) + \partial_{\varphi} \,\hat{\xi} (\varphi)\Big] = \gamma \,\hat{n} (\varphi)\;,
\end{equation}
where $\hat{\xi}(\varphi) = -\hat{a}_{r}(R,\varphi)$ becomes just a memory of the higher dimensional space and is absorbed in a new gauge potential, yielding Eq. \eqref{eq:kink1d} as a reduced expression for flux attachment. A similar logic can be applied in dimensionally reducing the parent gauge action in Eq. \eqref{eq:topo_action}. This is written schematically as $S_{2+1} \rightarrow S_{1+1}$, with identification
\begin{flalign}\label{eq:dim_red}
&S_{\text{BF}}\,[A, a] +  S_{\text{CS}}\,[A] \longrightarrow  S_{\text{Ax.}}\,[\Phi, a] + S_{\chi}\,[\Phi] \\ \label{eq:two_chiral}
&	\mathcal{L}_{1+1} =  \frac{\kappa}{2\pi}\, \Phi \,\epsilon^{\,\mu \nu}\partial_{\,\mu}\,a_{\nu} + \frac{\kappa}{2\pi}\, \epsilon^{\,0 1}\partial_{0}\,\Phi\,\partial_{1}\,\Phi  \;,
\end{flalign}
where $\mu, \nu =\{t, x\}\,$. We notice that $\mathcal{L}_{\text{Ax.}}$ in Eq. \eqref{eq:dim_red} constitutes both an axion and a BF term. It can also be understood, upon integration by parts, as a many-body Aharonov-Bohm twist effect \cite{santos2014manyab}. The axion contribution alone leads to a decoupling from the gauge field, but the introduction of $\mathcal{L}_{\chi}$ provides the axion with chiral dynamics. As expected, these contributions give a vanishing Hamiltonian, since they are first order in time derivatives, and are universal in that they do not depend on a specific matter model. Rather, they introduce constraints on the system, and fix the form of the statistical gauge field to be linear in density.

\paragraph*{\textit{Microscopic Emergence of a Statistical Gauge Field.}}
The previous discussion gives a macroscopic or phenomenological description of matter. However, it does not provide an intuitive explanation for how these statistical gauge fields could effectively be generated. It is evident from Eq. \eqref{eq:parity} that the minimal coupling to a linear-in-density gauge field [Eq. \eqref{eq:dens-dep_gauge}] can be seen as a combination of two and three-body contact interactions and an exotic parity-breaking term. The latter is a chiral interaction $\mathcal{H}_{\,\text{int}}\sim \,\colon(\mathbf{k}\,\hat{n}^{2}) \colon$ since it is momentum dependent in $\mathbf{k}$-space. We wonder whether it can arise from more conventional interaction terms. Hence, we consider a bosonic field theory with two-body interactions and a pseudo-spin $1/2$ degree of freedom $\alpha = \{\uparrow, \downarrow \}$. We focus only on central, spin-preserving interactions
\begin{equation}\label{eq:interact_term}
\sum_{\alpha,\beta} \int \text{d}\bm{x}\,\text{d}\bm{x'}\,\hat{\Psi}^{\dagger}_{\alpha}(\mathbf{x}) \,\hat{\Psi}^{\dagger}_{\beta}(\mathbf{x'})\,U^{\alpha \beta }(  \mathbf{x} - \mathbf{x'} )\, \hat{\Psi}_{\beta}(\mathbf{x'})\,\hat{\Psi}_{\alpha}(\mathbf{x}) \;.
\end{equation}
Upon rotation on the Bloch sphere and Fourier transforming we obtain the effective interaction in a new spin basis $\sigma = \{+, -\}$,  as
\begin{equation}
\sum_{\sigma,\tau,\sigma',\tau'} \,\int_{\mathbf{k}_{1},\mathbf{k}_{2},\mathbf{q}}\tilde{\chi}^{\,\sigma \tau \sigma' \tau'}_{\,\mathbf{k}_{1},\mathbf{k}_{2},\mathbf{q}}\,\hat{b}^{\dagger}_{\mathbf{k}_{1}+\mathbf{q},\sigma}\, \hat{b}^{\dagger}_{\mathbf{k}_{2}-\mathbf{q},\tau}\,\hat{b}_{\mathbf{k}_{2},\sigma'}\,\hat{b}_{\mathbf{k}_{1},\tau'} \;.
\end{equation}
We introduce $\int_{\mathbf{k}} \equiv \frac{1}{\sqrt{\text{vol.}}} \int \text{d}^{d}\mathbf{k} \,$ and define the screening function in momentum space as
\begin{equation}
	\tilde{\chi}^{\,\sigma \tau \sigma' \tau'}_{\,\mathbf{k}_{1},\mathbf{k}_{2},\mathbf{q}} = \sum_{\alpha,\beta} \tilde{U}^{\,\alpha \beta }_{\mathbf{q}}\; \hat{\eta}_{\,\mathbf{k}_{1}+\mathbf{q}}^{\dagger\;\alpha \sigma} \,\hat{\eta}_{\,\mathbf{k}_{2}-\mathbf{q}}^{\dagger\;\beta \tau} \,\hat{\eta}_{\,\mathbf{k}_{2}}^{\,\tau' \beta} \,\hat{\eta}_{\,\mathbf{k}_{1}}^{\,\sigma' \alpha} \;.
\end{equation}
Function $\tilde{\chi}$ can be interpreted as describing dressed interactions, while $\tilde{U}$ describes bare ones. The change of spin basis is nothing but a qubit  $\mathbf{k}-$dependent rotation parametrised by
\begin{equation}
	\hat{\eta}_{\,\mathbf{k}}^{\,\sigma \alpha} = \hat{\mathcal{R}}_{\,\hat{\mathbf{n}}_{\sigma}}(\theta_{\mathbf{k}} ) = \exp \Big( -i \,\theta_{\mathbf{k}} \,\frac{\hat{\mathbf{n}}_{\sigma} \cdot \vec{\bm{\sigma}}}{2} \Big)\,.
\end{equation}
A particular instance of the above scheme has been realised in a Raman-coupled Bose-Einstein condensate with internal atomic structure \cite{frolian2022realising,chisholm2022encoding}. As an instructive example, we consider the lowest-energy spin branch ($\sigma = \tau = \sigma' = \tau'$) and disregard the rest  ($\sigma \neq \tau \neq \sigma' \neq \tau'$). In this context, $(\pm)$ states are dressed states. We consider ultra-short-range bare interactions $U^{\alpha \beta }\,(\mathbf{x}) = g^{\alpha \beta}\,\delta\,(\mathbf{x})\,$, where $\delta\,(\mathbf{x})$ is the Dirac delta function. At low orders in $\mathbf{k} \approx \mathbf{k}_{0} + \delta\mathbf{k}$ expansion, with $\delta\mathbf{k} \ll 1$, we verify that the effective interaction kernel acquires the form
\begin{equation}
\tilde{\chi}^{\,\sigma \sigma \sigma \sigma}_{\,\mathbf{k}_{1}, \mathbf{k}_{2}, \mathbf{q}} \sim \mathcal{O}(\delta\mathbf{k}_{i}^{0})+ \mathcal{O}(\delta\mathbf{k}_{i}^{1}) \sim  (1+ \delta\mathbf{k}_{i})\,\delta_{\delta\mathbf{k}_{1} + \delta\mathbf{k}_{2}, \delta\mathbf{q}}\;.
\end{equation}
Thus, emergent longer range interactions now appear as a consequence of atomic light-dressing \cite{williams2012synthetic} and allow for the generation of the chiral interaction. 

Alternatively, one can view the density dependent gauge field arise in the mean field limit if a Pauli coupling term $\propto \hbar \Omega \, \vec{\bm{\sigma}}\cdot \vec{\bm{n}}$ is present in the Hamiltonian.
For weak enough particle interactions we can expand in a perturbation series and project onto one of the eigenstates of the system relying on the adiabatic theorem. The result is an effective Hamiltonian with emergent Berry connection terms arising perturbatively. The first order contribution $\bm{\mathcal{A}}^{(1)}_{\sigma} \propto \lvert\Psi_{\sigma}\rvert^{2}$ is linear in mean field density and can be computed in closed form. Hence, topological terms and the corresponding statistical gauge fields, like the density-dependent one discussed here \cite{edmonds2013simulating} or Chern-Simons \cite{valenti20synthetic} are not fictitious in the sense of a convenient computational trick, but can, instead, dynamically arise from effective microscopic interparticle interactions or as interacting Berry phase effects. 
\paragraph*{\textit{Lattice Version.}}
The previous ideas work equally well on the lattice, where they manifest as dynamical complex tunneling rates, being the general prescription of the form
\begin{equation}
\hat{H} = -J \sum_{j, \,\mu=1}^{d} \Big(\hat{c}^{\dagger}_{j} e^{\,i \hat{a}_{\mu} (j)} \hat{c}_{j+\hat{\text{e}}_{\mu}} + \text{H.c.}\Big) + \hat{H}_{\text{\,int}}
\end{equation}
where interactions are local, $\hat{c}$ is a bosonic (or fermionic) annihilation operator, $j$ denotes the lattice site and $\mu$ sums over nearest neighbours in $d$ spatial dimensions. Here, the statistical gauge field
\begin{equation}
\hat{a}_{\mu}\, (j) \equiv \hat{a}\,(j; j+\hat{\text{e}}_{\mu}) = \frac{1}{\hbar} \int^{\bm{x}_{j}}_{\bm{x}_{j}+\hat{\text{e}}_{j}} \text{d}\bm{x}\cdot\hat{\bm{a}}\,(t,\bm{x})
\end{equation}
 is the operator-valued Peierls phase
, defined on the links of the lattice, and accumulated in a tunneling event. The explicit form will, once again, depend on dimensionality, being $\hat{a}_{x} (j) = \gamma \,\hat{n}_{j}$ in $d=1$. In $d=2$ it is the solution, in any suitable gauge, of $\big(\hat{b}_{j} - \gamma\,\hat{n}_{j} \big) = 0$ with $\hat{b}_{j} = \epsilon^{ab} \Delta_{a}\, \hat{a}_{b}(j)$ being the ``magnetic" field associated with plaquette $j$ and $\Delta_{a}$ the lattice derivative in the $a^{\text{th}}$ direction. In $d=3$ it is the solution of  $\big(\Delta_{a} \,\hat{b}^{\,a}_{j} - \gamma\,\hat{n}_{j} \big) = 0\,$. It is worth observing that matter fields (i.e. electric charges) live on the direct lattice, while fluxes (i.e. magnetic charges) live on the dual lattice. Also, notice that recovering the continuum limit must be considered at all orders \cite{bonkhoff2020bosonic} in order to preserve compactness of the statistical parameter $\gamma \in [0,2\pi)$. Now, for the particular case of Hubbard-like interactions for $\hat{H}_{\text{int}}\,$, such a model reduces to an anyon-Hubbard model \cite{keilmann2011statistically} and has been extensively studied in 1d \cite{greschner2015hubbard}, but remains largely unexplored in other dimensions.

\paragraph*{\textit{Conclusions.}}


The composite particle duality extends the notion of flux attachment and survives in both lower and higher dimensions, and so does the picture of anyons understood as composites, although not necessarily as point-like objects. We support the assertion made by several authors \cite{hasenfratz1979puzzling, fradkin2017disorder, marino2017quantum} of an underlying \textit{order-disorder} scheme in Bose-Fermi correspondences based, according to our results, on the appearance of a statistical gauge fields, whose form is fixed by a topological field, leading to an effective transmutation mechanism. This is not only a physical process but it might also be experimentally controllable. The whole picture can also be observed through the lens of gauge transformations with topological features. In fact, we find large gauge transformations that coincide with a generalised Jordan-Wigner transformation. Our results remain valid either in continuum or the lattice, and for charged bosons or fermions as bare fields.
This work also opens the door to the experimental realisation of the aforementioned dualities, a range of anyonic models, and a new class of topological gauge theories that have not yet been studied in the context of quantum simulators, with the recent exception of \cite{frolian2022realising}. Yet, density-dependent gauge potentials have already been experimentally implemented \cite{frolian2022realising,clark2018density,gorg2019realization,lienhard2020soc,yao2022domain} in ultracold atomic platforms, so we are in hopes that such realisations can come to fruition in the near future.

\paragraph*{\textit{Acknowledgments.}}

We warmly thank C. Oliver, G. Palumbo, L. Tarruell for useful discussions. G.V-R. acknowledges financial support from EPSRC CM-CDT Grant No. EP/L015110/1. A.C. acknowledges financial support from MCIN/AEI/10.13039/501100011033 (LIGAS
PID2020-112687GB-C22) and from the Universitat Autònoma de Barcelona Talent Research program.

\bibliography{bibliotopo,compositeparticleNotes}

\begin{thebibliography}{48}%
\makeatletter
\providecommand \@ifxundefined [1]{%
 \@ifx{#1\undefined}
}%
\providecommand \@ifnum [1]{%
 \ifnum #1\expandafter \@firstoftwo
 \else \expandafter \@secondoftwo
 \fi
}%
\providecommand \@ifx [1]{%
 \ifx #1\expandafter \@firstoftwo
 \else \expandafter \@secondoftwo
 \fi
}%
\providecommand \natexlab [1]{#1}%
\providecommand \enquote  [1]{``#1''}%
\providecommand \bibnamefont  [1]{#1}%
\providecommand \bibfnamefont [1]{#1}%
\providecommand \citenamefont [1]{#1}%
\providecommand \href@noop [0]{\@secondoftwo}%
\providecommand \href [0]{\begingroup \@sanitize@url \@href}%
\providecommand \@href[1]{\@@startlink{#1}\@@href}%
\providecommand \@@href[1]{\endgroup#1\@@endlink}%
\providecommand \@sanitize@url [0]{\catcode `\\12\catcode `\$12\catcode
  `\&12\catcode `\#12\catcode `\^12\catcode `\_12\catcode `\%12\relax}%
\providecommand \@@startlink[1]{}%
\providecommand \@@endlink[0]{}%
\providecommand \url  [0]{\begingroup\@sanitize@url \@url }%
\providecommand \@url [1]{\endgroup\@href {#1}{\urlprefix }}%
\providecommand \urlprefix  [0]{URL }%
\providecommand \Eprint [0]{\href }%
\providecommand \doibase [0]{https://doi.org/}%
\providecommand \selectlanguage [0]{\@gobble}%
\providecommand \bibinfo  [0]{\@secondoftwo}%
\providecommand \bibfield  [0]{\@secondoftwo}%
\providecommand \translation [1]{[#1]}%
\providecommand \BibitemOpen [0]{}%
\providecommand \bibitemStop [0]{}%
\providecommand \bibitemNoStop [0]{.\EOS\space}%
\providecommand \EOS [0]{\spacefactor3000\relax}%
\providecommand \BibitemShut  [1]{\csname bibitem#1\endcsname}%
\let\auto@bib@innerbib\@empty
\bibitem [{\citenamefont {Aglietti}\ \emph {et~al.}(1996)\citenamefont
  {Aglietti}, \citenamefont {Griguolo}, \citenamefont {Jackiw}, \citenamefont
  {Pi},\ and\ \citenamefont {Seminara}}]{aglietti1996suplem}%
  \BibitemOpen
  \bibfield  {author} {\bibinfo {author} {\bibfnamefont {U.}~\bibnamefont
  {Aglietti}}, \bibinfo {author} {\bibfnamefont {L.}~\bibnamefont {Griguolo}},
  \bibinfo {author} {\bibfnamefont {R.}~\bibnamefont {Jackiw}}, \bibinfo
  {author} {\bibfnamefont {S.-Y.}\ \bibnamefont {Pi}},\ and\ \bibinfo {author}
  {\bibfnamefont {D.}~\bibnamefont {Seminara}},\ }\href
  {https://doi.org/10.1103/PhysRevLett.77.4406} {\bibfield  {journal} {\bibinfo
   {journal} {Phys. Rev. Lett.}\ }\textbf {\bibinfo {volume} {77}},\ \bibinfo
  {pages} {4406} (\bibinfo {year} {1996})}\BibitemShut {NoStop}%
\bibitem [{\citenamefont {Boyanovsky}\ \emph {et~al.}(1992)\citenamefont
  {Boyanovsky}, \citenamefont {Newman},\ and\ \citenamefont
  {Rovelli}}]{boyanovski1992cs}%
  \BibitemOpen
  \bibfield  {author} {\bibinfo {author} {\bibfnamefont {D.}~\bibnamefont
  {Boyanovsky}}, \bibinfo {author} {\bibfnamefont {E.~T.}\ \bibnamefont
  {Newman}},\ and\ \bibinfo {author} {\bibfnamefont {C.}~\bibnamefont
  {Rovelli}},\ }\href {https://doi.org/10.1103/PhysRevD.45.1210} {\bibfield
  {journal} {\bibinfo  {journal} {Phys. Rev. D}\ }\textbf {\bibinfo {volume}
  {45}},\ \bibinfo {pages} {1210} (\bibinfo {year} {1992})}\BibitemShut
  {NoStop}%
\bibitem [{\citenamefont {Benetton~Rabello}(1996)}]{rabello1996prl}%
  \BibitemOpen
  \bibfield  {author} {\bibinfo {author} {\bibfnamefont {S.~J.}\ \bibnamefont
  {Benetton~Rabello}},\ }\href {https://doi.org/10.1103/PhysRevLett.76.4007}
  {\bibfield  {journal} {\bibinfo  {journal} {Phys. Rev. Lett.}\ }\textbf
  {\bibinfo {volume} {76}},\ \bibinfo {pages} {4007} (\bibinfo {year}
  {1996})}\BibitemShut {NoStop}%
\bibitem [{\citenamefont {Rabello}(1995)}]{rabello1995gauge}%
  \BibitemOpen
  \bibfield  {author} {\bibinfo {author} {\bibfnamefont {S.~J.}\ \bibnamefont
  {Rabello}},\ }\href {https://doi.org/10.1016/0370-2693(95)01262-O} {\bibfield
   {journal} {\bibinfo  {journal} {Physics Letters B}\ }\textbf {\bibinfo
  {volume} {363}},\ \bibinfo {pages} {180} (\bibinfo {year}
  {1995})}\BibitemShut {NoStop}%
\bibitem [{\citenamefont {Jackiw}(1997)}]{jackiw1997review}%
  \BibitemOpen
  \bibfield  {author} {\bibinfo {author} {\bibfnamefont {R.}~\bibnamefont
  {Jackiw}},\ }\href {https://doi.org/10.2991/jnmp.1997.4.3-4.2} {\bibfield
  {journal} {\bibinfo  {journal} {Journal of Nonlinear Mathematical Physics}\
  }\textbf {\bibinfo {volume} {4}},\ \bibinfo {pages} {261} (\bibinfo {year}
  {1997})},\ \Eprint
  {https://arxiv.org/abs/https://doi.org/10.2991/jnmp.1997.4.3-4.2}
  {https://doi.org/10.2991/jnmp.1997.4.3-4.2} \BibitemShut {NoStop}%
\bibitem [{\citenamefont {Griguolo}\ and\ \citenamefont
  {Seminara}(1998)}]{griguolo1998chiral}%
  \BibitemOpen
  \bibfield  {author} {\bibinfo {author} {\bibfnamefont {L.}~\bibnamefont
  {Griguolo}}\ and\ \bibinfo {author} {\bibfnamefont {D.}~\bibnamefont
  {Seminara}},\ }\href {https://doi.org/10.1016/S0550-3213(97)00810-9}
  {\bibfield  {journal} {\bibinfo  {journal} {Nuclear Physics B}\ }\textbf
  {\bibinfo {volume} {516}},\ \bibinfo {pages} {467} (\bibinfo {year}
  {1998})}\BibitemShut {NoStop}%
\bibitem [{\citenamefont {Kundu}(1999)}]{kundy99anyons}%
  \BibitemOpen
  \bibfield  {author} {\bibinfo {author} {\bibfnamefont {A.}~\bibnamefont
  {Kundu}},\ }\href {https://doi.org/10.1103/PhysRevLett.83.1275} {\bibfield
  {journal} {\bibinfo  {journal} {Phys. Rev. Lett.}\ }\textbf {\bibinfo
  {volume} {83}},\ \bibinfo {pages} {1275} (\bibinfo {year}
  {1999})}\BibitemShut {NoStop}%
\bibitem [{\citenamefont {von Westenholz}(1979)}]{westenholz1979current}%
  \BibitemOpen
  \bibfield  {author} {\bibinfo {author} {\bibfnamefont {C.}~\bibnamefont {von
  Westenholz}},\ }\href {http://www.numdam.org/item/AIHPA_1979__30_4_353_0/}
  {\bibfield  {journal} {\bibinfo  {journal} {Annales de l'I.H.P. Physique
  th\'eorique}\ }\textbf {\bibinfo {volume} {30}},\ \bibinfo {pages} {353}
  (\bibinfo {year} {1979})}\BibitemShut {NoStop}%
\bibitem [{\citenamefont {Bhattacharjee}\ \emph {et~al.}(2017)\citenamefont
  {Bhattacharjee}, \citenamefont {Mj},\ and\ \citenamefont
  {Bandyopadhyay}}]{bhattacharjee2017topology}%
  \BibitemOpen
  \bibfield  {author} {\bibinfo {author} {\bibfnamefont {S.~M.}\ \bibnamefont
  {Bhattacharjee}}, \bibinfo {author} {\bibfnamefont {M.}~\bibnamefont {Mj}},\
  and\ \bibinfo {author} {\bibfnamefont {A.}~\bibnamefont {Bandyopadhyay}},\
  }\href@noop {} {\emph {\bibinfo {title} {Topology and Condensed Matter
  Physics}}},\ Vol.~\bibinfo {volume} {19}\ (\bibinfo  {publisher} {Springer},\
  \bibinfo {year} {2017})\BibitemShut {NoStop}%
\bibitem [{\citenamefont {Nakahara}(2018)}]{nakahara2018geometry}%
  \BibitemOpen
  \bibfield  {author} {\bibinfo {author} {\bibfnamefont {M.}~\bibnamefont
  {Nakahara}},\ }\href@noop {} {\emph {\bibinfo {title} {Geometry, topology and
  physics}}}\ (\bibinfo  {publisher} {CRC press},\ \bibinfo {year}
  {2018})\BibitemShut {NoStop}%
\bibitem [{\citenamefont {Chan}\ \emph {et~al.}(2013)\citenamefont {Chan},
  \citenamefont {Hughes}, \citenamefont {Ryu},\ and\ \citenamefont
  {Fradkin}}]{chan13functional}%
  \BibitemOpen
  \bibfield  {author} {\bibinfo {author} {\bibfnamefont {A.}~\bibnamefont
  {Chan}}, \bibinfo {author} {\bibfnamefont {T.~L.}\ \bibnamefont {Hughes}},
  \bibinfo {author} {\bibfnamefont {S.}~\bibnamefont {Ryu}},\ and\ \bibinfo
  {author} {\bibfnamefont {E.}~\bibnamefont {Fradkin}},\ }\href
  {https://doi.org/10.1103/PhysRevB.87.085132} {\bibfield  {journal} {\bibinfo
  {journal} {Phys. Rev. B}\ }\textbf {\bibinfo {volume} {87}},\ \bibinfo
  {pages} {085132} (\bibinfo {year} {2013})}\BibitemShut {NoStop}%
\bibitem [{\citenamefont {Palumbo}\ and\ \citenamefont
  {Goldman}(2019)}]{palumbo2019tensorberry}%
  \BibitemOpen
  \bibfield  {author} {\bibinfo {author} {\bibfnamefont {G.}~\bibnamefont
  {Palumbo}}\ and\ \bibinfo {author} {\bibfnamefont {N.}~\bibnamefont
  {Goldman}},\ }\href {https://doi.org/10.1103/PhysRevB.99.045154} {\bibfield
  {journal} {\bibinfo  {journal} {Phys. Rev. B}\ }\textbf {\bibinfo {volume}
  {99}},\ \bibinfo {pages} {045154} (\bibinfo {year} {2019})}\BibitemShut
  {NoStop}%
\bibitem [{\citenamefont {Abanov}(2017)}]{abanov2017theta}%
  \BibitemOpen
  \bibfield  {author} {\bibinfo {author} {\bibfnamefont {A.}~\bibnamefont
  {Abanov}},\ }\bibinfo {title} {Topology, geometry and quantum interference in
  condensed matter physics},\ in\ \href
  {https://doi.org/10.1007/978-981-10-6841-6_12} {\emph {\bibinfo {booktitle}
  {Topology and Condensed Matter Physics}}},\ \bibinfo {editor} {edited by\
  \bibinfo {editor} {\bibfnamefont {S.~M.}\ \bibnamefont {Bhattacharjee}},
  \bibinfo {editor} {\bibfnamefont {M.}~\bibnamefont {Mj}},\ and\ \bibinfo
  {editor} {\bibfnamefont {A.}~\bibnamefont {Bandyopadhyay}}}\ (\bibinfo
  {publisher} {Springer Singapore},\ \bibinfo {address} {Singapore},\ \bibinfo
  {year} {2017})\ pp.\ \bibinfo {pages} {281--331}\BibitemShut {NoStop}%
\bibitem [{\citenamefont {Fradkin}(2017)}]{fradkin2017disup}%
  \BibitemOpen
  \bibfield  {author} {\bibinfo {author} {\bibfnamefont {E.}~\bibnamefont
  {Fradkin}},\ }\href {https://doi.org/10.1007/s10955-017-1737-7} {\bibfield
  {journal} {\bibinfo  {journal} {Journal of Statistical Physics}\ }\textbf
  {\bibinfo {volume} {167}},\ \bibinfo {pages} {427} (\bibinfo {year}
  {2017})}\BibitemShut {NoStop}%
\bibitem [{\citenamefont {Fosco}\ and\ \citenamefont
  {Schaposnik}(2000)}]{fosco2000making}%
  \BibitemOpen
  \bibfield  {author} {\bibinfo {author} {\bibfnamefont {C.}~\bibnamefont
  {Fosco}}\ and\ \bibinfo {author} {\bibfnamefont {F.}~\bibnamefont
  {Schaposnik}},\ }\href {https://doi.org/10.1016/S0370-2693(00)00179-9}
  {\bibfield  {journal} {\bibinfo  {journal} {Physics Letters B}\ }\textbf
  {\bibinfo {volume} {477}},\ \bibinfo {pages} {341} (\bibinfo {year}
  {2000})}\BibitemShut {NoStop}%
\bibitem [{\citenamefont {Henneaux}\ and\ \citenamefont
  {Teitelboim}(1986)}]{henneaux1986p}%
  \BibitemOpen
  \bibfield  {author} {\bibinfo {author} {\bibfnamefont {M.}~\bibnamefont
  {Henneaux}}\ and\ \bibinfo {author} {\bibfnamefont {C.}~\bibnamefont
  {Teitelboim}},\ }\href@noop {} {\bibfield  {journal} {\bibinfo  {journal}
  {Foundations of Physics}\ }\textbf {\bibinfo {volume} {16}},\ \bibinfo
  {pages} {593} (\bibinfo {year} {1986})}\BibitemShut {NoStop}%
\bibitem [{\citenamefont {Wen}(1992)}]{wen1992theory}%
  \BibitemOpen
  \bibfield  {author} {\bibinfo {author} {\bibfnamefont {X.-G.}\ \bibnamefont
  {Wen}},\ }\href {https://doi.org/10.1142/S0217979292000840} {\bibfield
  {journal} {\bibinfo  {journal} {International journal of modern physics B}\
  }\textbf {\bibinfo {volume} {6}},\ \bibinfo {pages} {1711} (\bibinfo {year}
  {1992})}\BibitemShut {NoStop}%
\bibitem [{\citenamefont {Elitzur}\ \emph {et~al.}(1989)\citenamefont
  {Elitzur}, \citenamefont {Moore}, \citenamefont {Schwimmer},\ and\
  \citenamefont {Seiberg}}]{elitzur1989108}%
  \BibitemOpen
  \bibfield  {author} {\bibinfo {author} {\bibfnamefont {S.}~\bibnamefont
  {Elitzur}}, \bibinfo {author} {\bibfnamefont {G.}~\bibnamefont {Moore}},
  \bibinfo {author} {\bibfnamefont {A.}~\bibnamefont {Schwimmer}},\ and\
  \bibinfo {author} {\bibfnamefont {N.}~\bibnamefont {Seiberg}},\ }\href
  {https://doi.org/https://doi.org/10.1016/0550-3213(89)90436-7} {\bibfield
  {journal} {\bibinfo  {journal} {Nuclear Physics B}\ }\textbf {\bibinfo
  {volume} {326}},\ \bibinfo {pages} {108} (\bibinfo {year}
  {1989})}\BibitemShut {NoStop}%
\bibitem [{\citenamefont {Witten}(1989)}]{witten1989quantum}%
  \BibitemOpen
  \bibfield  {author} {\bibinfo {author} {\bibfnamefont {E.}~\bibnamefont
  {Witten}},\ }\href {https://doi.org/10.1007/BF01217730} {\bibfield  {journal}
  {\bibinfo  {journal} {Communications in Mathematical Physics}\ }\textbf
  {\bibinfo {volume} {121}},\ \bibinfo {pages} {351} (\bibinfo {year}
  {1989})}\BibitemShut {NoStop}%
\bibitem [{\citenamefont {Wen}(1990)}]{wen1990chiral}%
  \BibitemOpen
  \bibfield  {author} {\bibinfo {author} {\bibfnamefont {X.~G.}\ \bibnamefont
  {Wen}},\ }\href {https://doi.org/10.1103/PhysRevB.41.12838} {\bibfield
  {journal} {\bibinfo  {journal} {Phys. Rev. B}\ }\textbf {\bibinfo {volume}
  {41}},\ \bibinfo {pages} {12838} (\bibinfo {year} {1990})}\BibitemShut
  {NoStop}%
\bibitem [{\citenamefont {Jackiw}\ \emph {et~al.}(2000)\citenamefont {Jackiw},
  \citenamefont {Nair},\ and\ \citenamefont {Pi}}]{jackiw2000reduction}%
  \BibitemOpen
  \bibfield  {author} {\bibinfo {author} {\bibfnamefont {R.}~\bibnamefont
  {Jackiw}}, \bibinfo {author} {\bibfnamefont {V.~P.}\ \bibnamefont {Nair}},\
  and\ \bibinfo {author} {\bibfnamefont {S.-Y.}\ \bibnamefont {Pi}},\ }\href
  {https://doi.org/10.1103/PhysRevD.62.085018} {\bibfield  {journal} {\bibinfo
  {journal} {Phys. Rev. D}\ }\textbf {\bibinfo {volume} {62}},\ \bibinfo
  {pages} {085018} (\bibinfo {year} {2000})}\BibitemShut {NoStop}%
\bibitem [{\citenamefont {Qi}\ \emph {et~al.}(2008)\citenamefont {Qi},
  \citenamefont {Hughes},\ and\ \citenamefont {Zhang}}]{Qi2008tft}%
  \BibitemOpen
  \bibfield  {author} {\bibinfo {author} {\bibfnamefont {X.-L.}\ \bibnamefont
  {Qi}}, \bibinfo {author} {\bibfnamefont {T.~L.}\ \bibnamefont {Hughes}},\
  and\ \bibinfo {author} {\bibfnamefont {S.-C.}\ \bibnamefont {Zhang}},\ }\href
  {https://doi.org/10.1103/PhysRevB.78.195424} {\bibfield  {journal} {\bibinfo
  {journal} {Phys. Rev. B}\ }\textbf {\bibinfo {volume} {78}},\ \bibinfo
  {pages} {195424} (\bibinfo {year} {2008})}\BibitemShut {NoStop}%
\bibitem [{\citenamefont {Karch}\ \emph {et~al.}(2018)\citenamefont {Karch},
  \citenamefont {Tong},\ and\ \citenamefont {Turner}}]{Karch2018MirrorSA}%
  \BibitemOpen
  \bibfield  {author} {\bibinfo {author} {\bibfnamefont {A.}~\bibnamefont
  {Karch}}, \bibinfo {author} {\bibfnamefont {D.}~\bibnamefont {Tong}},\ and\
  \bibinfo {author} {\bibfnamefont {C.}~\bibnamefont {Turner}},\ }\href
  {https://doi.org/10.1007/JHEP07(2018)059} {\bibfield  {journal} {\bibinfo
  {journal} {Journal of High Energy Physics}\ }\textbf {\bibinfo {volume}
  {2018}},\ \bibinfo {pages} {1} (\bibinfo {year} {2018})}\BibitemShut
  {NoStop}%
\bibitem [{\citenamefont {Fradkin}(2013)}]{fradkin2013}%
  \BibitemOpen
  \bibfield  {author} {\bibinfo {author} {\bibfnamefont {E.}~\bibnamefont
  {Fradkin}},\ }\href {https://doi.org/10.1017/CBO9781139015509} {\emph
  {\bibinfo {title} {Field Theories of Condensed Matter Physics}}},\ \bibinfo
  {edition} {2nd}\ ed.\ (\bibinfo  {publisher} {Cambridge University Press},\
  \bibinfo {year} {2013})\BibitemShut {NoStop}%
\bibitem [{\citenamefont {Gaikwad}\ \emph {et~al.}(2020)\citenamefont
  {Gaikwad}, \citenamefont {Joshi}, \citenamefont {Mandal},\ and\ \citenamefont
  {Wadia}}]{gaikwad2020holographic}%
  \BibitemOpen
  \bibfield  {author} {\bibinfo {author} {\bibfnamefont {A.}~\bibnamefont
  {Gaikwad}}, \bibinfo {author} {\bibfnamefont {L.~K.}\ \bibnamefont {Joshi}},
  \bibinfo {author} {\bibfnamefont {G.}~\bibnamefont {Mandal}},\ and\ \bibinfo
  {author} {\bibfnamefont {S.~R.}\ \bibnamefont {Wadia}},\ }\href
  {https://doi.org/10.1007/JHEP02(2020)033} {\bibfield  {journal} {\bibinfo
  {journal} {Journal of High Energy Physics}\ }\textbf {\bibinfo {volume}
  {2020}},\ \bibinfo {pages} {1} (\bibinfo {year} {2020})}\BibitemShut
  {NoStop}%
\bibitem [{\citenamefont {Fosco}\ and\ \citenamefont
  {Schaposnik}(2020)}]{fosco2020d}%
  \BibitemOpen
  \bibfield  {author} {\bibinfo {author} {\bibfnamefont {C.}~\bibnamefont
  {Fosco}}\ and\ \bibinfo {author} {\bibfnamefont {F.}~\bibnamefont
  {Schaposnik}},\ }\href@noop {} {\bibfield  {journal} {\bibinfo  {journal}
  {arXiv preprint arXiv:2007.12993}\ } (\bibinfo {year} {2020})}\BibitemShut
  {NoStop}%
\bibitem [{\citenamefont {Harada}(1990)}]{harada90chiral}%
  \BibitemOpen
  \bibfield  {author} {\bibinfo {author} {\bibfnamefont {K.}~\bibnamefont
  {Harada}},\ }\href {https://doi.org/10.1103/PhysRevLett.64.139} {\bibfield
  {journal} {\bibinfo  {journal} {Phys. Rev. Lett.}\ }\textbf {\bibinfo
  {volume} {64}},\ \bibinfo {pages} {139} (\bibinfo {year} {1990})}\BibitemShut
  {NoStop}%
\bibitem [{\citenamefont {Absar}\ \emph {et~al.}()\citenamefont {Absar},
  \citenamefont {Ghoshal},\ and\ \citenamefont {Rahaman}}]{absarmodel}%
  \BibitemOpen
  \bibfield  {author} {\bibinfo {author} {\bibfnamefont {S.}~\bibnamefont
  {Absar}}, \bibinfo {author} {\bibfnamefont {S.}~\bibnamefont {Ghoshal}},\
  and\ \bibinfo {author} {\bibfnamefont {A.}~\bibnamefont {Rahaman}},\ }\href
  {https://doi.org/doi.org/10.1007/s10773-020-04680-1} {\bibinfo  {journal}
  {International Journal of Theoretical Physics}\ ,\ \bibinfo {pages}
  {1}}\BibitemShut {NoStop}%
\bibitem [{\citenamefont {Fr\"ohlich}\ and\ \citenamefont
  {Studer}(1993)}]{frohlich93review}%
  \BibitemOpen
\bibfield  {journal} {  }\bibfield  {author} {\bibinfo {author} {\bibfnamefont
  {J.}~\bibnamefont {Fr\"ohlich}}\ and\ \bibinfo {author} {\bibfnamefont
  {U.~M.}\ \bibnamefont {Studer}},\ }\href
  {https://doi.org/10.1103/RevModPhys.65.733} {\bibfield  {journal} {\bibinfo
  {journal} {Rev. Mod. Phys.}\ }\textbf {\bibinfo {volume} {65}},\ \bibinfo
  {pages} {733} (\bibinfo {year} {1993})}\BibitemShut {NoStop}%
\bibitem [{\citenamefont {Fr{\"o}hlich}\ and\ \citenamefont
  {Zee}(1991)}]{frohlich1991large}%
  \BibitemOpen
  \bibfield  {author} {\bibinfo {author} {\bibfnamefont {J.}~\bibnamefont
  {Fr{\"o}hlich}}\ and\ \bibinfo {author} {\bibfnamefont {A.}~\bibnamefont
  {Zee}},\ }\href {https://doi.org/10.1016/0550-3213(91)90275-3} {\bibfield
  {journal} {\bibinfo  {journal} {Nuclear Physics B}\ }\textbf {\bibinfo
  {volume} {364}},\ \bibinfo {pages} {517} (\bibinfo {year}
  {1991})}\BibitemShut {NoStop}%
\bibitem [{\citenamefont {Fr{\"o}hlich}\ and\ \citenamefont
  {Kerler}(1991)}]{frohlich1991universality}%
  \BibitemOpen
  \bibfield  {author} {\bibinfo {author} {\bibfnamefont {J.}~\bibnamefont
  {Fr{\"o}hlich}}\ and\ \bibinfo {author} {\bibfnamefont {T.}~\bibnamefont
  {Kerler}},\ }\href {https://doi.org/10.1016/0550-3213(91)90360-A} {\bibfield
  {journal} {\bibinfo  {journal} {Nuclear Physics B}\ }\textbf {\bibinfo
  {volume} {354}},\ \bibinfo {pages} {369} (\bibinfo {year}
  {1991})}\BibitemShut {NoStop}%
\bibitem [{\citenamefont {Marchetti}\ \emph {et~al.}(1996)\citenamefont
  {Marchetti}, \citenamefont {Su},\ and\ \citenamefont
  {Yu}}]{marchetti1996dimensional}%
  \BibitemOpen
  \bibfield  {author} {\bibinfo {author} {\bibfnamefont {P.}~\bibnamefont
  {Marchetti}}, \bibinfo {author} {\bibfnamefont {Z.-B.}\ \bibnamefont {Su}},\
  and\ \bibinfo {author} {\bibfnamefont {L.}~\bibnamefont {Yu}},\ }\href
  {https://doi.org/10.1016/S0550-3213(96)00458-0} {\bibfield  {journal}
  {\bibinfo  {journal} {Nuclear Physics B}\ }\textbf {\bibinfo {volume}
  {482}},\ \bibinfo {pages} {731} (\bibinfo {year} {1996})}\BibitemShut
  {NoStop}%
\bibitem [{\citenamefont {Karch}\ \emph {et~al.}(2019)\citenamefont {Karch},
  \citenamefont {Tong},\ and\ \citenamefont {Turner}}]{karch2019arf}%
  \BibitemOpen
  \bibfield  {author} {\bibinfo {author} {\bibfnamefont {A.}~\bibnamefont
  {Karch}}, \bibinfo {author} {\bibfnamefont {D.}~\bibnamefont {Tong}},\ and\
  \bibinfo {author} {\bibfnamefont {C.}~\bibnamefont {Turner}},\ }\href
  {https://doi.org/10.21468/SciPostPhys.7.1.007} {\bibfield  {journal}
  {\bibinfo  {journal} {SciPost Phys.}\ }\textbf {\bibinfo {volume} {7}},\
  \bibinfo {pages} {7} (\bibinfo {year} {2019})}\BibitemShut {NoStop}%
\bibitem [{\citenamefont {Okuda}\ \emph {et~al.}(2021)\citenamefont {Okuda},
  \citenamefont {Saito},\ and\ \citenamefont {Yokoyama}}]{okuda2021u}%
  \BibitemOpen
  \bibfield  {author} {\bibinfo {author} {\bibfnamefont {T.}~\bibnamefont
  {Okuda}}, \bibinfo {author} {\bibfnamefont {K.}~\bibnamefont {Saito}},\ and\
  \bibinfo {author} {\bibfnamefont {S.}~\bibnamefont {Yokoyama}},\ }\href
  {https://doi.org/10.1016/j.nuclphysb.2020.115272} {\bibfield  {journal}
  {\bibinfo  {journal} {Nuclear Physics B}\ }\textbf {\bibinfo {volume}
  {962}},\ \bibinfo {pages} {115272} (\bibinfo {year} {2021})}\BibitemShut
  {NoStop}%
\bibitem [{\citenamefont {Senthil}\ \emph {et~al.}(2019)\citenamefont
  {Senthil}, \citenamefont {Son}, \citenamefont {Wang},\ and\ \citenamefont
  {Xu}}]{senthil2019duality}%
  \BibitemOpen
  \bibfield  {author} {\bibinfo {author} {\bibfnamefont {T.}~\bibnamefont
  {Senthil}}, \bibinfo {author} {\bibfnamefont {D.~T.}\ \bibnamefont {Son}},
  \bibinfo {author} {\bibfnamefont {C.}~\bibnamefont {Wang}},\ and\ \bibinfo
  {author} {\bibfnamefont {C.}~\bibnamefont {Xu}},\ }\href
  {https://doi.org/10.1016/j.physrep.2019.09.001} {\bibfield  {journal}
  {\bibinfo  {journal} {Physics Reports}\ }\textbf {\bibinfo {volume} {827}},\
  \bibinfo {pages} {1} (\bibinfo {year} {2019})}\BibitemShut {NoStop}%
\bibitem [{\citenamefont {Giamarchi}(2003)}]{giamarchi2003quantum}%
  \BibitemOpen
  \bibfield  {author} {\bibinfo {author} {\bibfnamefont {T.}~\bibnamefont
  {Giamarchi}},\ }\href@noop {} {\emph {\bibinfo {title} {Quantum physics in
  one dimension}}},\ Vol.\ \bibinfo {volume} {121}\ (\bibinfo  {publisher}
  {Clarendon press},\ \bibinfo {year} {2003})\BibitemShut {NoStop}%
\bibitem [{\citenamefont {Keilmann}\ \emph {et~al.}(2011)\citenamefont
  {Keilmann}, \citenamefont {Lanzmich}, \citenamefont {McCulloch},\ and\
  \citenamefont {Roncaglia}}]{keilmann2011statistically}%
  \BibitemOpen
  \bibfield  {author} {\bibinfo {author} {\bibfnamefont {T.}~\bibnamefont
  {Keilmann}}, \bibinfo {author} {\bibfnamefont {S.}~\bibnamefont {Lanzmich}},
  \bibinfo {author} {\bibfnamefont {I.}~\bibnamefont {McCulloch}},\ and\
  \bibinfo {author} {\bibfnamefont {M.}~\bibnamefont {Roncaglia}},\ }\href
  {https://doi.org/10.1038/ncomms1353} {\bibfield  {journal} {\bibinfo
  {journal} {Nature communications}\ }\textbf {\bibinfo {volume} {2}},\
  \bibinfo {pages} {1} (\bibinfo {year} {2011})}\BibitemShut {NoStop}%
\bibitem [{\citenamefont {Bonkhoff}\ \emph {et~al.}(2020)\citenamefont
  {Bonkhoff}, \citenamefont {J{\"a}gering}, \citenamefont {Eggert},
  \citenamefont {Pelster}, \citenamefont {Thorwart},\ and\ \citenamefont
  {Posske}}]{bonkhoff2020bosonic}%
  \BibitemOpen
  \bibfield  {author} {\bibinfo {author} {\bibfnamefont {M.}~\bibnamefont
  {Bonkhoff}}, \bibinfo {author} {\bibfnamefont {K.}~\bibnamefont
  {J{\"a}gering}}, \bibinfo {author} {\bibfnamefont {S.}~\bibnamefont
  {Eggert}}, \bibinfo {author} {\bibfnamefont {A.}~\bibnamefont {Pelster}},
  \bibinfo {author} {\bibfnamefont {M.}~\bibnamefont {Thorwart}},\ and\
  \bibinfo {author} {\bibfnamefont {T.}~\bibnamefont {Posske}},\ }\href@noop {}
  {\bibfield  {journal} {\bibinfo  {journal} {arXiv preprint arXiv:2008.00003}\
  } (\bibinfo {year} {2020})}\BibitemShut {NoStop}%
\bibitem [{\citenamefont {Hirsch}(1989)}]{hirsch1989bond}%
  \BibitemOpen
  \bibfield  {author} {\bibinfo {author} {\bibfnamefont {J.}~\bibnamefont
  {Hirsch}},\ }\href {https://doi.org/10.1016/0921-4534(89)90225-6} {\bibfield
  {journal} {\bibinfo  {journal} {Physica C: Superconductivity and its
  Applications}\ }\textbf {\bibinfo {volume} {158}},\ \bibinfo {pages} {326}
  (\bibinfo {year} {1989})}\BibitemShut {NoStop}%
\bibitem [{\citenamefont {Mark}\ and\ \citenamefont
  {Motrunich}(2020)}]{mark2020eta}%
  \BibitemOpen
  \bibfield  {author} {\bibinfo {author} {\bibfnamefont {D.~K.}\ \bibnamefont
  {Mark}}\ and\ \bibinfo {author} {\bibfnamefont {O.~I.}\ \bibnamefont
  {Motrunich}},\ }\href {https://doi.org/10.1103/PhysRevB.102.075132}
  {\bibfield  {journal} {\bibinfo  {journal} {Phys. Rev. B}\ }\textbf {\bibinfo
  {volume} {102}},\ \bibinfo {pages} {075132} (\bibinfo {year}
  {2020})}\BibitemShut {NoStop}%
\bibitem [{\citenamefont {Vitoriano}\ and\ \citenamefont
  {Coutinho-Filho}(2009)}]{carlindo2009bond}%
  \BibitemOpen
  \bibfield  {author} {\bibinfo {author} {\bibfnamefont {C.}~\bibnamefont
  {Vitoriano}}\ and\ \bibinfo {author} {\bibfnamefont {M.~D.}\ \bibnamefont
  {Coutinho-Filho}},\ }\href {https://doi.org/10.1103/PhysRevLett.102.146404}
  {\bibfield  {journal} {\bibinfo  {journal} {Phys. Rev. Lett.}\ }\textbf
  {\bibinfo {volume} {102}},\ \bibinfo {pages} {146404} (\bibinfo {year}
  {2009})}\BibitemShut {NoStop}%
\bibitem [{\citenamefont {Anfossi}\ \emph {et~al.}(2006)\citenamefont
  {Anfossi}, \citenamefont {Boschi}, \citenamefont {Montorsi},\ and\
  \citenamefont {Ortolani}}]{anfossi2006entangle}%
  \BibitemOpen
  \bibfield  {author} {\bibinfo {author} {\bibfnamefont {A.}~\bibnamefont
  {Anfossi}}, \bibinfo {author} {\bibfnamefont {C.~D.~E.}\ \bibnamefont
  {Boschi}}, \bibinfo {author} {\bibfnamefont {A.}~\bibnamefont {Montorsi}},\
  and\ \bibinfo {author} {\bibfnamefont {F.}~\bibnamefont {Ortolani}},\ }\href
  {https://doi.org/10.1103/PhysRevB.73.085113} {\bibfield  {journal} {\bibinfo
  {journal} {Phys. Rev. B}\ }\textbf {\bibinfo {volume} {73}},\ \bibinfo
  {pages} {085113} (\bibinfo {year} {2006})}\BibitemShut {NoStop}%
\bibitem [{\citenamefont {Dutta}\ \emph {et~al.}(2015)\citenamefont {Dutta},
  \citenamefont {Gajda}, \citenamefont {Hauke}, \citenamefont {Lewenstein},
  \citenamefont {Lühmann}, \citenamefont {Malomed}, \citenamefont
  {Sowi{\'{n}}ski},\ and\ \citenamefont {Zakrzewski}}]{Dutta_2015}%
  \BibitemOpen
  \bibfield  {author} {\bibinfo {author} {\bibfnamefont {O.}~\bibnamefont
  {Dutta}}, \bibinfo {author} {\bibfnamefont {M.}~\bibnamefont {Gajda}},
  \bibinfo {author} {\bibfnamefont {P.}~\bibnamefont {Hauke}}, \bibinfo
  {author} {\bibfnamefont {M.}~\bibnamefont {Lewenstein}}, \bibinfo {author}
  {\bibfnamefont {D.-S.}\ \bibnamefont {Lühmann}}, \bibinfo {author}
  {\bibfnamefont {B.~A.}\ \bibnamefont {Malomed}}, \bibinfo {author}
  {\bibfnamefont {T.}~\bibnamefont {Sowi{\'{n}}ski}},\ and\ \bibinfo {author}
  {\bibfnamefont {J.}~\bibnamefont {Zakrzewski}},\ }\href
  {https://doi.org/10.1088/0034-4885/78/6/066001} {\bibfield  {journal}
  {\bibinfo  {journal} {Reports on Progress in Physics}\ }\textbf {\bibinfo
  {volume} {78}},\ \bibinfo {pages} {066001} (\bibinfo {year}
  {2015})}\BibitemShut {NoStop}%
\bibitem [{\citenamefont {Duan}(2007)}]{Duan_2007}%
  \BibitemOpen
  \bibfield  {author} {\bibinfo {author} {\bibfnamefont {L.-M.}\ \bibnamefont
  {Duan}},\ }\href {https://doi.org/10.1209/0295-5075/81/20001} {\bibfield
  {journal} {\bibinfo  {journal} {{EPL} (Europhysics Letters)}\ }\textbf
  {\bibinfo {volume} {81}},\ \bibinfo {pages} {20001} (\bibinfo {year}
  {2007})}\BibitemShut {NoStop}%
\bibitem [{\citenamefont {Ezawa}(2013)}]{ezawa2013quantum}%
  \BibitemOpen
  \bibfield  {author} {\bibinfo {author} {\bibfnamefont {Z.~F.}\ \bibnamefont
  {Ezawa}},\ }\href {https://doi.org/10.1142/8210} {\emph {\bibinfo {title}
  {Quantum Hall effects: Recent theoretical and experimental developments}}},\
  \bibinfo {edition} {3rd}\ ed.\ (\bibinfo  {publisher} {World Scientific
  Publishing Company},\ \bibinfo {year} {2013})\BibitemShut {NoStop}%
\bibitem [{\citenamefont {Batchelor}\ \emph {et~al.}(2008)\citenamefont
  {Batchelor}, \citenamefont {Guan},\ and\ \citenamefont
  {Kundu}}]{Batchelor2008}%
  \BibitemOpen
  \bibfield  {author} {\bibinfo {author} {\bibfnamefont {M.~T.}\ \bibnamefont
  {Batchelor}}, \bibinfo {author} {\bibfnamefont {X.-W.}\ \bibnamefont
  {Guan}},\ and\ \bibinfo {author} {\bibfnamefont {A.}~\bibnamefont {Kundu}},\
  }\href {https://doi.org/10.1088/1751-8113/41/35/352002} {\bibfield  {journal}
  {\bibinfo  {journal} {Journal of Physics A: Mathematical and Theoretical}\
  }\textbf {\bibinfo {volume} {41}},\ \bibinfo {pages} {352002} (\bibinfo
  {year} {2008})}\BibitemShut {NoStop}%
\bibitem [{\citenamefont {Bellazzini}(2010)}]{Bellazzini2010}%
  \BibitemOpen
  \bibfield  {author} {\bibinfo {author} {\bibfnamefont {B.}~\bibnamefont
  {Bellazzini}},\ }\href {https://doi.org/10.1088/1751-8113/44/3/035403}
  {\bibfield  {journal} {\bibinfo  {journal} {Journal of Physics A:
  Mathematical and Theoretical}\ }\textbf {\bibinfo {volume} {44}},\ \bibinfo
  {pages} {035403} (\bibinfo {year} {2010})}\BibitemShut {NoStop}%
\bibitem [{\citenamefont {Patton}\ and\ \citenamefont
  {Sheehy}(2020)}]{patton2020hfb}%
  \BibitemOpen
  \bibfield  {author} {\bibinfo {author} {\bibfnamefont {K.~R.}\ \bibnamefont
  {Patton}}\ and\ \bibinfo {author} {\bibfnamefont {D.~E.}\ \bibnamefont
  {Sheehy}},\ }\href {https://doi.org/10.1103/PhysRevA.101.063607} {\bibfield
  {journal} {\bibinfo  {journal} {Phys. Rev. A}\ }\textbf {\bibinfo {volume}
  {101}},\ \bibinfo {pages} {063607} (\bibinfo {year} {2020})}\BibitemShut
  {NoStop}%
\end{thebibliography}%


\begin{thebibliography}{58}%
\makeatletter
\providecommand \@ifxundefined [1]{%
 \@ifx{#1\undefined}
}%
\providecommand \@ifnum [1]{%
 \ifnum #1\expandafter \@firstoftwo
 \else \expandafter \@secondoftwo
 \fi
}%
\providecommand \@ifx [1]{%
 \ifx #1\expandafter \@firstoftwo
 \else \expandafter \@secondoftwo
 \fi
}%
\providecommand \natexlab [1]{#1}%
\providecommand \enquote  [1]{``#1''}%
\providecommand \bibnamefont  [1]{#1}%
\providecommand \bibfnamefont [1]{#1}%
\providecommand \citenamefont [1]{#1}%
\providecommand \href@noop [0]{\@secondoftwo}%
\providecommand \href [0]{\begingroup \@sanitize@url \@href}%
\providecommand \@href[1]{\@@startlink{#1}\@@href}%
\providecommand \@@href[1]{\endgroup#1\@@endlink}%
\providecommand \@sanitize@url [0]{\catcode `\\12\catcode `\$12\catcode
  `\&12\catcode `\#12\catcode `\^12\catcode `\_12\catcode `\%12\relax}%
\providecommand \@@startlink[1]{}%
\providecommand \@@endlink[0]{}%
\providecommand \url  [0]{\begingroup\@sanitize@url \@url }%
\providecommand \@url [1]{\endgroup\@href {#1}{\urlprefix }}%
\providecommand \urlprefix  [0]{URL }%
\providecommand \Eprint [0]{\href }%
\providecommand \doibase [0]{https://doi.org/}%
\providecommand \selectlanguage [0]{\@gobble}%
\providecommand \bibinfo  [0]{\@secondoftwo}%
\providecommand \bibfield  [0]{\@secondoftwo}%
\providecommand \translation [1]{[#1]}%
\providecommand \BibitemOpen [0]{}%
\providecommand \bibitemStop [0]{}%
\providecommand \bibitemNoStop [0]{.\EOS\space}%
\providecommand \EOS [0]{\spacefactor3000\relax}%
\providecommand \BibitemShut  [1]{\csname bibitem#1\endcsname}%
\let\auto@bib@innerbib\@empty
\bibitem [{\citenamefont {Wilczek}(1982{\natexlab{a}})}]{wilckez1982flux}%
  \BibitemOpen
  \bibfield  {author} {\bibinfo {author} {\bibfnamefont {F.}~\bibnamefont
  {Wilczek}},\ }\bibfield  {title} {\bibinfo {title} {Magnetic flux, angular
  momentum, and statistics},\ }\href
  {https://doi.org/10.1103/PhysRevLett.48.1144} {\bibfield  {journal} {\bibinfo
   {journal} {Phys. Rev. Lett.}\ }\textbf {\bibinfo {volume} {48}},\ \bibinfo
  {pages} {1144} (\bibinfo {year} {1982}{\natexlab{a}})}\BibitemShut {NoStop}%
\bibitem [{\citenamefont {Wilczek}(1982{\natexlab{b}})}]{wilczek1982fraction}%
  \BibitemOpen
  \bibfield  {author} {\bibinfo {author} {\bibfnamefont {F.}~\bibnamefont
  {Wilczek}},\ }\bibfield  {title} {\bibinfo {title} {Quantum mechanics of
  fractional-spin particles},\ }\href
  {https://doi.org/10.1103/PhysRevLett.49.957} {\bibfield  {journal} {\bibinfo
  {journal} {Phys. Rev. Lett.}\ }\textbf {\bibinfo {volume} {49}},\ \bibinfo
  {pages} {957} (\bibinfo {year} {1982}{\natexlab{b}})}\BibitemShut {NoStop}%
\bibitem [{\citenamefont {Wen}(1990)}]{wen1990topological}%
  \BibitemOpen
  \bibfield  {author} {\bibinfo {author} {\bibfnamefont {X.-G.}\ \bibnamefont
  {Wen}},\ }\bibfield  {title} {\bibinfo {title} {Topological orders in rigid
  states},\ }\href {https://doi.org/10.1142/S0217979290000139} {\bibfield
  {journal} {\bibinfo  {journal} {International Journal of Modern Physics B}\
  }\textbf {\bibinfo {volume} {4}},\ \bibinfo {pages} {239} (\bibinfo {year}
  {1990})}\BibitemShut {NoStop}%
\bibitem [{\citenamefont {Moessner}\ and\ \citenamefont
  {Moore}(2021)}]{moessner2021topological}%
  \BibitemOpen
  \bibfield  {author} {\bibinfo {author} {\bibfnamefont {R.}~\bibnamefont
  {Moessner}}\ and\ \bibinfo {author} {\bibfnamefont {J.~E.}\ \bibnamefont
  {Moore}},\ }\href@noop {} {\emph {\bibinfo {title} {Topological phases of
  matter}}}\ (\bibinfo  {publisher} {Cambridge University Press},\ \bibinfo
  {year} {2021})\BibitemShut {NoStop}%
\bibitem [{\citenamefont {Feldman}\ and\ \citenamefont
  {Halperin}(2021)}]{feldman2021fractional}%
  \BibitemOpen
  \bibfield  {author} {\bibinfo {author} {\bibfnamefont {D.~E.}\ \bibnamefont
  {Feldman}}\ and\ \bibinfo {author} {\bibfnamefont {B.~I.}\ \bibnamefont
  {Halperin}},\ }\bibfield  {title} {\bibinfo {title} {Fractional charge and
  fractional statistics in the quantum hall effects},\ }\href
  {https://doi.org/10.1088/1361-6633/ac03aa} {\bibfield  {journal} {\bibinfo
  {journal} {Reports on Progress in Physics}\ }\textbf {\bibinfo {volume}
  {84}},\ \bibinfo {pages} {076501} (\bibinfo {year} {2021})}\BibitemShut
  {NoStop}%
\bibitem [{\citenamefont {Forte}(1992)}]{forte1992review}%
  \BibitemOpen
  \bibfield  {author} {\bibinfo {author} {\bibfnamefont {S.}~\bibnamefont
  {Forte}},\ }\bibfield  {title} {\bibinfo {title} {Quantum mechanics and field
  theory with fractional spin and statistics},\ }\href
  {https://doi.org/10.1103/RevModPhys.64.193} {\bibfield  {journal} {\bibinfo
  {journal} {Rev. Mod. Phys.}\ }\textbf {\bibinfo {volume} {64}},\ \bibinfo
  {pages} {193} (\bibinfo {year} {1992})}\BibitemShut {NoStop}%
\bibitem [{\citenamefont {Polyakov}(1988)}]{polyakov1988fermi}%
  \BibitemOpen
  \bibfield  {author} {\bibinfo {author} {\bibfnamefont {A.~M.}\ \bibnamefont
  {Polyakov}},\ }\bibfield  {title} {\bibinfo {title} {Fermi-bose
  transmutations induced by gauge fields},\ }\href
  {https://doi.org/10.1142/S0217732388000398} {\bibfield  {journal} {\bibinfo
  {journal} {Modern Physics Letters A}\ }\textbf {\bibinfo {volume} {3}},\
  \bibinfo {pages} {325} (\bibinfo {year} {1988})}\BibitemShut {NoStop}%
\bibitem [{\citenamefont {Leinaas}\ and\ \citenamefont
  {Myrheim}(1977)}]{leinaas1977theory}%
  \BibitemOpen
  \bibfield  {author} {\bibinfo {author} {\bibfnamefont {J.~M.}\ \bibnamefont
  {Leinaas}}\ and\ \bibinfo {author} {\bibfnamefont {J.}~\bibnamefont
  {Myrheim}},\ }\bibfield  {title} {\bibinfo {title} {On the theory of
  identical particles},\ }\href {https://doi.org/10.1007/BF02727953} {\bibfield
   {journal} {\bibinfo  {journal} {Il Nuovo Cimento B (1971-1996)}\ }\textbf
  {\bibinfo {volume} {37}},\ \bibinfo {pages} {1} (\bibinfo {year}
  {1977})}\BibitemShut {NoStop}%
\bibitem [{\citenamefont {Jordan}\ and\ \citenamefont
  {Wigner}(1928)}]{jordan1928paulische}%
  \BibitemOpen
  \bibfield  {author} {\bibinfo {author} {\bibfnamefont {P.}~\bibnamefont
  {Jordan}}\ and\ \bibinfo {author} {\bibfnamefont {E.}~\bibnamefont
  {Wigner}},\ }\bibfield  {title} {\bibinfo {title} {{\"U}ber das paulische
  {\"a}quivalenzverbot},\ }\href {https://doi.org/10.1007/BF01331938}
  {\bibfield  {journal} {\bibinfo  {journal} {Zeitschrift f\"ur Physik}\
  }\textbf {\bibinfo {volume} {47}},\ \bibinfo {pages} {631} (\bibinfo {year}
  {1928})}\BibitemShut {NoStop}%
\bibitem [{\citenamefont {Coleman}(1975)}]{coleman1975sine}%
  \BibitemOpen
  \bibfield  {author} {\bibinfo {author} {\bibfnamefont {S.}~\bibnamefont
  {Coleman}},\ }\bibfield  {title} {\bibinfo {title} {Quantum sine-gordon
  equation as the massive thirring model},\ }\href
  {https://doi.org/10.1103/PhysRevD.11.2088} {\bibfield  {journal} {\bibinfo
  {journal} {Phys. Rev. D}\ }\textbf {\bibinfo {volume} {11}},\ \bibinfo
  {pages} {2088} (\bibinfo {year} {1975})}\BibitemShut {NoStop}%
\bibitem [{\citenamefont {Mandelstam}(1975)}]{mandelstam1975soliton}%
  \BibitemOpen
  \bibfield  {author} {\bibinfo {author} {\bibfnamefont {S.}~\bibnamefont
  {Mandelstam}},\ }\bibfield  {title} {\bibinfo {title} {Soliton operators for
  the quantized sine-gordon equation},\ }\href
  {https://doi.org/10.1103/PhysRevD.11.3026} {\bibfield  {journal} {\bibinfo
  {journal} {Phys. Rev. D}\ }\textbf {\bibinfo {volume} {11}},\ \bibinfo
  {pages} {3026} (\bibinfo {year} {1975})}\BibitemShut {NoStop}%
\bibitem [{\citenamefont {Valiente}(2021)}]{valiente2021universal}%
  \BibitemOpen
  \bibfield  {author} {\bibinfo {author} {\bibfnamefont {M.}~\bibnamefont
  {Valiente}},\ }\bibfield  {title} {\bibinfo {title} {Universal duality
  transformations in interacting one-dimensional quantum systems},\ }\href
  {https://doi.org/10.1103/PhysRevA.103.L021302} {\bibfield  {journal}
  {\bibinfo  {journal} {Phys. Rev. A}\ }\textbf {\bibinfo {volume} {103}},\
  \bibinfo {pages} {L021302} (\bibinfo {year} {2021})}\BibitemShut {NoStop}%
\bibitem [{\citenamefont {Harshman}\ and\ \citenamefont
  {Knapp}(2022)}]{Harshman2022exchange}%
  \BibitemOpen
  \bibfield  {author} {\bibinfo {author} {\bibfnamefont {N.~L.}\ \bibnamefont
  {Harshman}}\ and\ \bibinfo {author} {\bibfnamefont {A.~C.}\ \bibnamefont
  {Knapp}},\ }\bibfield  {title} {\bibinfo {title} {Topological exchange
  statistics in one dimension},\ }\href
  {https://doi.org/10.1103/PhysRevA.105.052214} {\bibfield  {journal} {\bibinfo
   {journal} {Phys. Rev. A}\ }\textbf {\bibinfo {volume} {105}},\ \bibinfo
  {pages} {052214} (\bibinfo {year} {2022})}\BibitemShut {NoStop}%
\bibitem [{\citenamefont {Benetton~Rabello}(1996)}]{rabello1996prl}%
  \BibitemOpen
  \bibfield  {author} {\bibinfo {author} {\bibfnamefont {S.~J.}\ \bibnamefont
  {Benetton~Rabello}},\ }\bibfield  {title} {\bibinfo {title} {1d generalized
  statistics gas: A gauge theory approach},\ }\href
  {https://doi.org/10.1103/PhysRevLett.76.4007} {\bibfield  {journal} {\bibinfo
   {journal} {Phys. Rev. Lett.}\ }\textbf {\bibinfo {volume} {76}},\ \bibinfo
  {pages} {4007} (\bibinfo {year} {1996})}\BibitemShut {NoStop}%
\bibitem [{\citenamefont {Aglietti}\ \emph {et~al.}(1996)\citenamefont
  {Aglietti}, \citenamefont {Griguolo}, \citenamefont {Jackiw}, \citenamefont
  {Pi},\ and\ \citenamefont {Seminara}}]{aglietti1996solitons}%
  \BibitemOpen
  \bibfield  {author} {\bibinfo {author} {\bibfnamefont {U.}~\bibnamefont
  {Aglietti}}, \bibinfo {author} {\bibfnamefont {L.}~\bibnamefont {Griguolo}},
  \bibinfo {author} {\bibfnamefont {R.}~\bibnamefont {Jackiw}}, \bibinfo
  {author} {\bibfnamefont {S.-Y.}\ \bibnamefont {Pi}},\ and\ \bibinfo {author}
  {\bibfnamefont {D.}~\bibnamefont {Seminara}},\ }\bibfield  {title} {\bibinfo
  {title} {Anyons and chiral solitons on a line},\ }\href
  {https://doi.org/10.1103/PhysRevLett.77.4406} {\bibfield  {journal} {\bibinfo
   {journal} {Phys. Rev. Lett.}\ }\textbf {\bibinfo {volume} {77}},\ \bibinfo
  {pages} {4406} (\bibinfo {year} {1996})}\BibitemShut {NoStop}%
\bibitem [{\citenamefont {Kundu}(1999)}]{kundy99anyons}%
  \BibitemOpen
  \bibfield  {author} {\bibinfo {author} {\bibfnamefont {A.}~\bibnamefont
  {Kundu}},\ }\bibfield  {title} {\bibinfo {title} {Exact solution of double
  $\ensuremath{\delta}$ function bose gas through an interacting anyon gas},\
  }\href {https://doi.org/10.1103/PhysRevLett.83.1275} {\bibfield  {journal}
  {\bibinfo  {journal} {Phys. Rev. Lett.}\ }\textbf {\bibinfo {volume} {83}},\
  \bibinfo {pages} {1275} (\bibinfo {year} {1999})}\BibitemShut {NoStop}%
\bibitem [{\citenamefont {Fr{\"o}lian}\ \emph {et~al.}(2022)\citenamefont
  {Fr{\"o}lian}, \citenamefont {Chisholm}, \citenamefont {Neri}, \citenamefont
  {Cabrera}, \citenamefont {Ramos}, \citenamefont {Celi},\ and\ \citenamefont
  {Tarruell}}]{frolian2022realising}%
  \BibitemOpen
  \bibfield  {author} {\bibinfo {author} {\bibfnamefont {A.}~\bibnamefont
  {Fr{\"o}lian}}, \bibinfo {author} {\bibfnamefont {C.~S.}\ \bibnamefont
  {Chisholm}}, \bibinfo {author} {\bibfnamefont {E.}~\bibnamefont {Neri}},
  \bibinfo {author} {\bibfnamefont {C.~R.}\ \bibnamefont {Cabrera}}, \bibinfo
  {author} {\bibfnamefont {R.}~\bibnamefont {Ramos}}, \bibinfo {author}
  {\bibfnamefont {A.}~\bibnamefont {Celi}},\ and\ \bibinfo {author}
  {\bibfnamefont {L.}~\bibnamefont {Tarruell}},\ }\bibfield  {title} {\bibinfo
  {title} {Realizing a 1d topological gauge theory in an optically dressed
  bec},\ }\href {https://doi.org/10.1038/s41586-022-04943-3} {\bibfield
  {journal} {\bibinfo  {journal} {Nature}\ }\textbf {\bibinfo {volume} {608}},\
  \bibinfo {pages} {293} (\bibinfo {year} {2022})}\BibitemShut {NoStop}%
\bibitem [{\citenamefont {Chisholm}\ \emph {et~al.}(2022)\citenamefont
  {Chisholm}, \citenamefont {Fr{\"o}lian}, \citenamefont {Neri}, \citenamefont
  {Ramos}, \citenamefont {Tarruell},\ and\ \citenamefont
  {Celi}}]{chisholm2022encoding}%
  \BibitemOpen
  \bibfield  {author} {\bibinfo {author} {\bibfnamefont {C.~S.}\ \bibnamefont
  {Chisholm}}, \bibinfo {author} {\bibfnamefont {A.}~\bibnamefont
  {Fr{\"o}lian}}, \bibinfo {author} {\bibfnamefont {E.}~\bibnamefont {Neri}},
  \bibinfo {author} {\bibfnamefont {R.}~\bibnamefont {Ramos}}, \bibinfo
  {author} {\bibfnamefont {L.}~\bibnamefont {Tarruell}},\ and\ \bibinfo
  {author} {\bibfnamefont {A.}~\bibnamefont {Celi}},\ }\bibfield  {title}
  {\bibinfo {title} {Encoding a one-dimensional topological gauge theory in a
  raman-coupled bose-einstein condensate},\ }\href
  {https://doi.org/10.1103/PhysRevResearch.4.043088} {\bibfield  {journal}
  {\bibinfo  {journal} {Phys. Rev. Research}\ }\textbf {\bibinfo {volume}
  {4}},\ \bibinfo {pages} {043088} (\bibinfo {year} {2022})}\BibitemShut
  {NoStop}%
\bibitem [{\citenamefont {Marino}(2017)}]{marino2017quantum}%
  \BibitemOpen
  \bibfield  {author} {\bibinfo {author} {\bibfnamefont {E.~C.}\ \bibnamefont
  {Marino}},\ }\href {https://doi.org/10.1017/9781139696548} {\emph {\bibinfo
  {title} {Quantum field theory approach to condensed matter physics}}}\
  (\bibinfo  {publisher} {Cambridge University Press},\ \bibinfo {year}
  {2017})\BibitemShut {NoStop}%
\bibitem [{\citenamefont {Wilczek}(1990)}]{wilczek1990book}%
  \BibitemOpen
  \bibfield  {author} {\bibinfo {author} {\bibfnamefont {F.}~\bibnamefont
  {Wilczek}},\ }\href@noop {} {\emph {\bibinfo {title} {Fractional statistics
  and anyon superconductivity}}},\ Vol.~\bibinfo {volume} {5}\ (\bibinfo
  {publisher} {World scientific},\ \bibinfo {year} {1990})\BibitemShut
  {NoStop}%
\bibitem [{\citenamefont {Kadanoff}\ and\ \citenamefont
  {Ceva}(1971)}]{kadanoffceva1971}%
  \BibitemOpen
  \bibfield  {author} {\bibinfo {author} {\bibfnamefont {L.~P.}\ \bibnamefont
  {Kadanoff}}\ and\ \bibinfo {author} {\bibfnamefont {H.}~\bibnamefont
  {Ceva}},\ }\bibfield  {title} {\bibinfo {title} {Determination of an operator
  algebra for the two-dimensional ising model},\ }\href
  {https://doi.org/10.1103/PhysRevB.3.3918} {\bibfield  {journal} {\bibinfo
  {journal} {Phys. Rev. B}\ }\textbf {\bibinfo {volume} {3}},\ \bibinfo {pages}
  {3918} (\bibinfo {year} {1971})}\BibitemShut {NoStop}%
\bibitem [{\citenamefont {Fradkin}(2017)}]{fradkin2017disorder}%
  \BibitemOpen
  \bibfield  {author} {\bibinfo {author} {\bibfnamefont {E.}~\bibnamefont
  {Fradkin}},\ }\bibfield  {title} {\bibinfo {title} {Disorder operators and
  their descendants},\ }\href {https://doi.org/10.1007/s10955-017-1737-7}
  {\bibfield  {journal} {\bibinfo  {journal} {Journal of Statistical Physics}\
  }\textbf {\bibinfo {volume} {167}},\ \bibinfo {pages} {427} (\bibinfo {year}
  {2017})}\BibitemShut {NoStop}%
\bibitem [{Note1()}]{Note1}%
  \BibitemOpen
  \bibinfo {note} {This is meant to generalise the concept of a Jordan-Wigner
  string to higher dimensions. For low dimensions the \protect \textit {brane},
  and thus the composite operator, are local, i.e. they can be defined at a
  given point in space. However, for $\protect \text {D}\ge 3$ this is not the
  case and these objects become intrinsically non-local, a feature captured by
  $\Gamma _{\protect \mathbf {x}}$.}\BibitemShut {Stop}%
\bibitem [{\citenamefont {Le~Guillou}\ \emph {et~al.}(1996)\citenamefont
  {Le~Guillou}, \citenamefont {N{\'u}{\~n}ez},\ and\ \citenamefont
  {Schaposnik}}]{le1996current}%
  \BibitemOpen
  \bibfield  {author} {\bibinfo {author} {\bibfnamefont {J.~C.}\ \bibnamefont
  {Le~Guillou}}, \bibinfo {author} {\bibfnamefont {C.}~\bibnamefont
  {N{\'u}{\~n}ez}},\ and\ \bibinfo {author} {\bibfnamefont {F.~A.}\
  \bibnamefont {Schaposnik}},\ }\bibfield  {title} {\bibinfo {title} {Current
  algebra and bosonization in three dimensions},\ }\href
  {https://doi.org/10.1006/aphy.1996.0120} {\bibfield  {journal} {\bibinfo
  {journal} {Annals of Physics}\ }\textbf {\bibinfo {volume} {251}},\ \bibinfo
  {pages} {426} (\bibinfo {year} {1996})}\BibitemShut {NoStop}%
\bibitem [{\citenamefont {Fradkin}\ and\ \citenamefont
  {Schaposnik}(1994)}]{fradkin1994fermion}%
  \BibitemOpen
  \bibfield  {author} {\bibinfo {author} {\bibfnamefont {E.}~\bibnamefont
  {Fradkin}}\ and\ \bibinfo {author} {\bibfnamefont {F.~A.}\ \bibnamefont
  {Schaposnik}},\ }\bibfield  {title} {\bibinfo {title} {The fermion-boson
  mapping in three-dimensional quantum field theory},\ }\href
  {https://doi.org/10.1016/0370-2693(94)91374-9} {\bibfield  {journal}
  {\bibinfo  {journal} {Physics Letters B}\ }\textbf {\bibinfo {volume}
  {338}},\ \bibinfo {pages} {253} (\bibinfo {year} {1994})}\BibitemShut
  {NoStop}%
\bibitem [{\citenamefont {Frohlich}\ \emph {et~al.}(1995)\citenamefont
  {Frohlich}, \citenamefont {Gotschmann},\ and\ \citenamefont
  {Marchetti}}]{frohlich1995bosonize}%
  \BibitemOpen
  \bibfield  {author} {\bibinfo {author} {\bibfnamefont {J.}~\bibnamefont
  {Frohlich}}, \bibinfo {author} {\bibfnamefont {R.}~\bibnamefont
  {Gotschmann}},\ and\ \bibinfo {author} {\bibfnamefont {P.~A.}\ \bibnamefont
  {Marchetti}},\ }\bibfield  {title} {\bibinfo {title} {Bosonization of fermi
  systems in arbitrary dimension in terms of gauge forms},\ }\href
  {https://doi.org/10.1088/0305-4470/28/5/008} {\bibfield  {journal} {\bibinfo
  {journal} {Journal of Physics A: Mathematical and General}\ }\textbf
  {\bibinfo {volume} {28}},\ \bibinfo {pages} {1169} (\bibinfo {year}
  {1995})}\BibitemShut {NoStop}%
\bibitem [{\citenamefont {Burgess}\ and\ \citenamefont
  {Quevedo}(1994)}]{burgess1994bosonization}%
  \BibitemOpen
  \bibfield  {author} {\bibinfo {author} {\bibfnamefont {C.}~\bibnamefont
  {Burgess}}\ and\ \bibinfo {author} {\bibfnamefont {F.}~\bibnamefont
  {Quevedo}},\ }\bibfield  {title} {\bibinfo {title} {Bosonization as
  duality},\ }\href {https://doi.org/10.1016/0550-3213(94)90332-8} {\bibfield
  {journal} {\bibinfo  {journal} {Nuclear Physics B}\ }\textbf {\bibinfo
  {volume} {421}},\ \bibinfo {pages} {373} (\bibinfo {year}
  {1994})}\BibitemShut {NoStop}%
\bibitem [{\citenamefont {Burgess}\ \emph {et~al.}(1994)\citenamefont
  {Burgess}, \citenamefont {L{\"u}tken},\ and\ \citenamefont
  {Quevedo}}]{burgess1994bosonization2}%
  \BibitemOpen
  \bibfield  {author} {\bibinfo {author} {\bibfnamefont {C.}~\bibnamefont
  {Burgess}}, \bibinfo {author} {\bibfnamefont {C.}~\bibnamefont
  {L{\"u}tken}},\ and\ \bibinfo {author} {\bibfnamefont {F.}~\bibnamefont
  {Quevedo}},\ }\bibfield  {title} {\bibinfo {title} {Bosonization in higher
  dimensions},\ }\href {https://doi.org/10.1016/0370-2693(94)00963-5}
  {\bibfield  {journal} {\bibinfo  {journal} {Physics Letters B}\ }\textbf
  {\bibinfo {volume} {336}},\ \bibinfo {pages} {18} (\bibinfo {year}
  {1994})}\BibitemShut {NoStop}%
\bibitem [{\citenamefont {Schaposnik}(1995)}]{schaposnik1995comment}%
  \BibitemOpen
  \bibfield  {author} {\bibinfo {author} {\bibfnamefont {F.}~\bibnamefont
  {Schaposnik}},\ }\bibfield  {title} {\bibinfo {title} {A comment on
  bosonization in {$d > 2$} dimensions},\ }\href
  {https://doi.org/10.1016/0370-2693(95)00776-H} {\bibfield  {journal}
  {\bibinfo  {journal} {Physics Letters B}\ }\textbf {\bibinfo {volume}
  {356}},\ \bibinfo {pages} {39} (\bibinfo {year} {1995})}\BibitemShut
  {NoStop}%
\bibitem [{\citenamefont {Fosco}\ and\ \citenamefont
  {Schaposnik}(2018)}]{fosco2018functional}%
  \BibitemOpen
  \bibfield  {author} {\bibinfo {author} {\bibfnamefont {C.~D.}\ \bibnamefont
  {Fosco}}\ and\ \bibinfo {author} {\bibfnamefont {F.~A.}\ \bibnamefont
  {Schaposnik}},\ }\bibfield  {title} {\bibinfo {title} {Functional
  bosonization of a dirac field in 2+1 dimensions, in the presence of a
  boundary},\ }\href {https://doi.org/10.1016/j.physletb.2018.05.032}
  {\bibfield  {journal} {\bibinfo  {journal} {Physics Letters B}\ }\textbf
  {\bibinfo {volume} {782}},\ \bibinfo {pages} {224} (\bibinfo {year}
  {2018})}\BibitemShut {NoStop}%
\bibitem [{\citenamefont {Schaposnik}(1998)}]{schaposnik1998bosonization}%
  \BibitemOpen
  \bibfield  {author} {\bibinfo {author} {\bibfnamefont {F.~A.}\ \bibnamefont
  {Schaposnik}},\ }\bibfield  {title} {\bibinfo {title} {Bosonization in {$d >
  2$} dimensions},\ }\href {https://doi.org/10.1063/1.54689} {\bibfield
  {journal} {\bibinfo  {journal} {AIP Conference Proceedings}\ }\textbf
  {\bibinfo {volume} {419}},\ \bibinfo {pages} {151} (\bibinfo {year}
  {1998})}\BibitemShut {NoStop}%
\bibitem [{\citenamefont {Banerjee}\ and\ \citenamefont
  {Marino}(1997)}]{banerjee1997501}%
  \BibitemOpen
  \bibfield  {author} {\bibinfo {author} {\bibfnamefont {R.}~\bibnamefont
  {Banerjee}}\ and\ \bibinfo {author} {\bibfnamefont {E.}~\bibnamefont
  {Marino}},\ }\bibfield  {title} {\bibinfo {title} {Different approaches for
  bosonization in higher dimensions},\ }\href
  {https://doi.org/https://doi.org/10.1016/S0550-3213(97)00549-X} {\bibfield
  {journal} {\bibinfo  {journal} {Nuclear Physics B}\ }\textbf {\bibinfo
  {volume} {507}},\ \bibinfo {pages} {501} (\bibinfo {year}
  {1997})}\BibitemShut {NoStop}%
\bibitem [{\citenamefont {Chan}\ \emph {et~al.}(2013)\citenamefont {Chan},
  \citenamefont {Hughes}, \citenamefont {Ryu},\ and\ \citenamefont
  {Fradkin}}]{chanmain13functional}%
  \BibitemOpen
  \bibfield  {author} {\bibinfo {author} {\bibfnamefont {A.}~\bibnamefont
  {Chan}}, \bibinfo {author} {\bibfnamefont {T.~L.}\ \bibnamefont {Hughes}},
  \bibinfo {author} {\bibfnamefont {S.}~\bibnamefont {Ryu}},\ and\ \bibinfo
  {author} {\bibfnamefont {E.}~\bibnamefont {Fradkin}},\ }\bibfield  {title}
  {\bibinfo {title} {Effective field theories for topological insulators by
  functional bosonization},\ }\href
  {https://doi.org/10.1103/PhysRevB.87.085132} {\bibfield  {journal} {\bibinfo
  {journal} {Phys. Rev. B}\ }\textbf {\bibinfo {volume} {87}},\ \bibinfo
  {pages} {085132} (\bibinfo {year} {2013})}\BibitemShut {NoStop}%
\bibitem [{\citenamefont {Cirio}\ \emph {et~al.}(2014)\citenamefont {Cirio},
  \citenamefont {Palumbo},\ and\ \citenamefont {Pachos}}]{cirio2014tight}%
  \BibitemOpen
  \bibfield  {author} {\bibinfo {author} {\bibfnamefont {M.}~\bibnamefont
  {Cirio}}, \bibinfo {author} {\bibfnamefont {G.}~\bibnamefont {Palumbo}},\
  and\ \bibinfo {author} {\bibfnamefont {J.~K.}\ \bibnamefont {Pachos}},\
  }\bibfield  {title} {\bibinfo {title} {$(3+1)$-dimensional topological
  quantum field theory from a tight-binding model of interacting spinless
  fermions},\ }\href {https://doi.org/10.1103/PhysRevB.90.085114} {\bibfield
  {journal} {\bibinfo  {journal} {Phys. Rev. B}\ }\textbf {\bibinfo {volume}
  {90}},\ \bibinfo {pages} {085114} (\bibinfo {year} {2014})}\BibitemShut
  {NoStop}%
\bibitem [{\citenamefont {Jackiw}\ \emph {et~al.}(2004)\citenamefont {Jackiw},
  \citenamefont {Nair}, \citenamefont {Pi},\ and\ \citenamefont
  {Polychronakos}}]{jackiw2004perfect}%
  \BibitemOpen
  \bibfield  {author} {\bibinfo {author} {\bibfnamefont {R.}~\bibnamefont
  {Jackiw}}, \bibinfo {author} {\bibfnamefont {V.}~\bibnamefont {Nair}},
  \bibinfo {author} {\bibfnamefont {S.}~\bibnamefont {Pi}},\ and\ \bibinfo
  {author} {\bibfnamefont {A.}~\bibnamefont {Polychronakos}},\ }\bibfield
  {title} {\bibinfo {title} {Perfect fluid theory and its extensions},\ }\href
  {https://doi.org/10.1088/0305-4470/37/42/R01} {\bibfield  {journal} {\bibinfo
   {journal} {Journal of Physics A: Mathematical and General}\ }\textbf
  {\bibinfo {volume} {37}},\ \bibinfo {pages} {R327} (\bibinfo {year}
  {2004})}\BibitemShut {NoStop}%
\bibitem [{Note2()}]{Note2}%
  \BibitemOpen
  \bibinfo {note} {We use the shorthand notation $\protect \mathbf {x}
  =(t,\protect \bm {x})$ unless explicitly noted otherwise and the convention
  that charge or number density is given by $\protect \hat {\protect \mathcal
  {J}}^{\protect \,0} = \protect \hat {n}$. In doing so we are taking the
  electric unit of charge $q=c=1$ to be equal to one.}\BibitemShut {Stop}%
\bibitem [{\citenamefont {Schwinger}(1969)}]{schwinger1969magnetic}%
  \BibitemOpen
  \bibfield  {author} {\bibinfo {author} {\bibfnamefont {J.}~\bibnamefont
  {Schwinger}},\ }\bibfield  {title} {\bibinfo {title} {A magnetic model of
  matter},\ }\href {https://doi.org/10.1126/science.165.3895.757} {\bibfield
  {journal} {\bibinfo  {journal} {Science}\ }\textbf {\bibinfo {volume}
  {165}},\ \bibinfo {pages} {757} (\bibinfo {year} {1969})}\BibitemShut
  {NoStop}%
\bibitem [{\citenamefont {Schwinger}(1968)}]{schwinger1968sources}%
  \BibitemOpen
  \bibfield  {author} {\bibinfo {author} {\bibfnamefont {J.}~\bibnamefont
  {Schwinger}},\ }\bibfield  {title} {\bibinfo {title} {Sources and magnetic
  charge},\ }\href {https://doi.org/10.1103/PhysRev.173.1536} {\bibfield
  {journal} {\bibinfo  {journal} {Phys. Rev.}\ }\textbf {\bibinfo {volume}
  {173}},\ \bibinfo {pages} {1536} (\bibinfo {year} {1968})}\BibitemShut
  {NoStop}%
\bibitem [{\citenamefont {Schwinger}(1966)}]{schwinger1966charge}%
  \BibitemOpen
  \bibfield  {author} {\bibinfo {author} {\bibfnamefont {J.}~\bibnamefont
  {Schwinger}},\ }\bibfield  {title} {\bibinfo {title} {Magnetic charge and
  quantum field theory},\ }\href {https://doi.org/10.1103/PhysRev.144.1087}
  {\bibfield  {journal} {\bibinfo  {journal} {Phys. Rev.}\ }\textbf {\bibinfo
  {volume} {144}},\ \bibinfo {pages} {1087} (\bibinfo {year}
  {1966})}\BibitemShut {NoStop}%
\bibitem [{Note3()}]{Note3}%
  \BibitemOpen
  \bibinfo {note} {We emphasise that the bare species chosen could also be
  fermionic and the mechanism would work analogously. Similarly, for a
  relativistic model such as a complex Klein-Gordon field, our claims still
  hold. Here we choose a bosonic field for convenience and experimental
  relevance in ultracold atoms.}\BibitemShut {Stop}%
\bibitem [{\citenamefont {Edmonds}\ \emph {et~al.}(2013)\citenamefont
  {Edmonds}, \citenamefont {Valiente}, \citenamefont
  {Juzeli\ifmmode~\bar{u}\else \={u}\fi{}nas}, \citenamefont {Santos},\ and\
  \citenamefont {\"Ohberg}}]{edmonds2013simulating}%
  \BibitemOpen
  \bibfield  {author} {\bibinfo {author} {\bibfnamefont {M.~J.}\ \bibnamefont
  {Edmonds}}, \bibinfo {author} {\bibfnamefont {M.}~\bibnamefont {Valiente}},
  \bibinfo {author} {\bibfnamefont {G.}~\bibnamefont
  {Juzeli\ifmmode~\bar{u}\else \={u}\fi{}nas}}, \bibinfo {author}
  {\bibfnamefont {L.}~\bibnamefont {Santos}},\ and\ \bibinfo {author}
  {\bibfnamefont {P.}~\bibnamefont {\"Ohberg}},\ }\bibfield  {title} {\bibinfo
  {title} {Simulating an interacting gauge theory with ultracold bose gases},\
  }\href {https://doi.org/10.1103/PhysRevLett.110.085301} {\bibfield  {journal}
  {\bibinfo  {journal} {Phys. Rev. Lett.}\ }\textbf {\bibinfo {volume} {110}},\
  \bibinfo {pages} {085301} (\bibinfo {year} {2013})}\BibitemShut {NoStop}%
\bibitem [{\citenamefont {Calabrese}\ and\ \citenamefont
  {Mintchev}(2007)}]{calabrese2007correlation}%
  \BibitemOpen
  \bibfield  {author} {\bibinfo {author} {\bibfnamefont {P.}~\bibnamefont
  {Calabrese}}\ and\ \bibinfo {author} {\bibfnamefont {M.}~\bibnamefont
  {Mintchev}},\ }\bibfield  {title} {\bibinfo {title} {Correlation functions of
  one-dimensional anyonic fluids},\ }\href
  {https://doi.org/10.1103/PhysRevB.75.233104} {\bibfield  {journal} {\bibinfo
  {journal} {Phys. Rev. B}\ }\textbf {\bibinfo {volume} {75}},\ \bibinfo
  {pages} {233104} (\bibinfo {year} {2007})}\BibitemShut {NoStop}%
\bibitem [{\citenamefont {Piroli}\ \emph {et~al.}(2020)\citenamefont {Piroli},
  \citenamefont {Scopa},\ and\ \citenamefont
  {Calabrese}}]{piroli2020determinant}%
  \BibitemOpen
  \bibfield  {author} {\bibinfo {author} {\bibfnamefont {L.}~\bibnamefont
  {Piroli}}, \bibinfo {author} {\bibfnamefont {S.}~\bibnamefont {Scopa}},\ and\
  \bibinfo {author} {\bibfnamefont {P.}~\bibnamefont {Calabrese}},\ }\bibfield
  {title} {\bibinfo {title} {Determinant formula for the field form factor in
  the anyonic lieb--liniger model},\ }\href
  {https://doi.org/10.1088/1751-8121/ab94ed} {\bibfield  {journal} {\bibinfo
  {journal} {Journal of Physics A: Mathematical and Theoretical}\ }\textbf
  {\bibinfo {volume} {53}},\ \bibinfo {pages} {405001} (\bibinfo {year}
  {2020})}\BibitemShut {NoStop}%
\bibitem [{\citenamefont {Scopa}\ \emph {et~al.}(2020)\citenamefont {Scopa},
  \citenamefont {Piroli},\ and\ \citenamefont {Calabrese}}]{scopa2020one}%
  \BibitemOpen
  \bibfield  {author} {\bibinfo {author} {\bibfnamefont {S.}~\bibnamefont
  {Scopa}}, \bibinfo {author} {\bibfnamefont {L.}~\bibnamefont {Piroli}},\ and\
  \bibinfo {author} {\bibfnamefont {P.}~\bibnamefont {Calabrese}},\ }\bibfield
  {title} {\bibinfo {title} {One-particle density matrix of a trapped
  lieb--liniger anyonic gas},\ }\href
  {https://doi.org/10.1088/1742-5468/abaed1} {\bibfield  {journal} {\bibinfo
  {journal} {Journal of Statistical Mechanics: Theory and Experiment}\ }\textbf
  {\bibinfo {volume} {2020}},\ \bibinfo {pages} {093103} (\bibinfo {year}
  {2020})}\BibitemShut {NoStop}%
\bibitem [{\citenamefont {Girardeau}(2006)}]{girardeau2006anyon}%
  \BibitemOpen
  \bibfield  {author} {\bibinfo {author} {\bibfnamefont {M.~D.}\ \bibnamefont
  {Girardeau}},\ }\bibfield  {title} {\bibinfo {title} {Anyon-fermion mapping
  and applications to ultracold gases in tight waveguides},\ }\href
  {https://doi.org/10.1103/PhysRevLett.97.100402} {\bibfield  {journal}
  {\bibinfo  {journal} {Phys. Rev. Lett.}\ }\textbf {\bibinfo {volume} {97}},\
  \bibinfo {pages} {100402} (\bibinfo {year} {2006})}\BibitemShut {NoStop}%
\bibitem [{\citenamefont {Girardeau}(1960)}]{girardeau1960relationship}%
  \BibitemOpen
  \bibfield  {author} {\bibinfo {author} {\bibfnamefont {M.}~\bibnamefont
  {Girardeau}},\ }\bibfield  {title} {\bibinfo {title} {Relationship between
  systems of impenetrable bosons and fermions in one dimension},\ }\href
  {https://doi.org/10.1063/1.1703687} {\bibfield  {journal} {\bibinfo
  {journal} {Journal of Mathematical Physics}\ }\textbf {\bibinfo {volume}
  {1}},\ \bibinfo {pages} {516} (\bibinfo {year} {1960})}\BibitemShut {NoStop}%
\bibitem [{\citenamefont {Cheon}\ and\ \citenamefont
  {Shigehara}(1999)}]{cheon1999fbduality}%
  \BibitemOpen
  \bibfield  {author} {\bibinfo {author} {\bibfnamefont {T.}~\bibnamefont
  {Cheon}}\ and\ \bibinfo {author} {\bibfnamefont {T.}~\bibnamefont
  {Shigehara}},\ }\bibfield  {title} {\bibinfo {title} {Fermion-boson duality
  of one-dimensional quantum particles with generalized contact interactions},\
  }\href {https://doi.org/10.1103/PhysRevLett.82.2536} {\bibfield  {journal}
  {\bibinfo  {journal} {Phys. Rev. Lett.}\ }\textbf {\bibinfo {volume} {82}},\
  \bibinfo {pages} {2536} (\bibinfo {year} {1999})}\BibitemShut {NoStop}%
\bibitem [{\citenamefont {Santos}\ and\ \citenamefont
  {Wang}(2014)}]{santos2014manyab}%
  \BibitemOpen
  \bibfield  {author} {\bibinfo {author} {\bibfnamefont {L.~H.}\ \bibnamefont
  {Santos}}\ and\ \bibinfo {author} {\bibfnamefont {J.}~\bibnamefont {Wang}},\
  }\bibfield  {title} {\bibinfo {title} {Symmetry-protected many-body
  aharonov-bohm effect},\ }\href {https://doi.org/10.1103/PhysRevB.89.195122}
  {\bibfield  {journal} {\bibinfo  {journal} {Phys. Rev. B}\ }\textbf {\bibinfo
  {volume} {89}},\ \bibinfo {pages} {195122} (\bibinfo {year}
  {2014})}\BibitemShut {NoStop}%
\bibitem [{\citenamefont {Williams}\ \emph {et~al.}(2012)\citenamefont
  {Williams}, \citenamefont {LeBlanc}, \citenamefont {Jimenez-Garcia},
  \citenamefont {Beeler}, \citenamefont {Perry}, \citenamefont {Phillips},\
  and\ \citenamefont {Spielman}}]{williams2012synthetic}%
  \BibitemOpen
  \bibfield  {author} {\bibinfo {author} {\bibfnamefont {R.~A.}\ \bibnamefont
  {Williams}}, \bibinfo {author} {\bibfnamefont {L.~J.}\ \bibnamefont
  {LeBlanc}}, \bibinfo {author} {\bibfnamefont {K.}~\bibnamefont
  {Jimenez-Garcia}}, \bibinfo {author} {\bibfnamefont {M.~C.}\ \bibnamefont
  {Beeler}}, \bibinfo {author} {\bibfnamefont {A.~R.}\ \bibnamefont {Perry}},
  \bibinfo {author} {\bibfnamefont {W.~D.}\ \bibnamefont {Phillips}},\ and\
  \bibinfo {author} {\bibfnamefont {I.~B.}\ \bibnamefont {Spielman}},\
  }\bibfield  {title} {\bibinfo {title} {Synthetic partial waves in ultracold
  atomic collisions},\ }\href {https://doi.org/0.1126/science.1212652}
  {\bibfield  {journal} {\bibinfo  {journal} {Science}\ }\textbf {\bibinfo
  {volume} {335}},\ \bibinfo {pages} {314} (\bibinfo {year}
  {2012})}\BibitemShut {NoStop}%
\bibitem [{\citenamefont {Valent\'{\i}-Rojas}\ \emph
  {et~al.}(2020)\citenamefont {Valent\'{\i}-Rojas}, \citenamefont
  {Westerberg},\ and\ \citenamefont {\"Ohberg}}]{valenti20synthetic}%
  \BibitemOpen
  \bibfield  {author} {\bibinfo {author} {\bibfnamefont {G.}~\bibnamefont
  {Valent\'{\i}-Rojas}}, \bibinfo {author} {\bibfnamefont {N.}~\bibnamefont
  {Westerberg}},\ and\ \bibinfo {author} {\bibfnamefont {P.}~\bibnamefont
  {\"Ohberg}},\ }\bibfield  {title} {\bibinfo {title} {Synthetic flux
  attachment},\ }\href {https://doi.org/10.1103/PhysRevResearch.2.033453}
  {\bibfield  {journal} {\bibinfo  {journal} {Phys. Rev. Research}\ }\textbf
  {\bibinfo {volume} {2}},\ \bibinfo {pages} {033453} (\bibinfo {year}
  {2020})}\BibitemShut {NoStop}%
\bibitem [{\citenamefont {Bonkhoff}\ \emph {et~al.}(2021)\citenamefont
  {Bonkhoff}, \citenamefont {J\"agering}, \citenamefont {Eggert}, \citenamefont
  {Pelster}, \citenamefont {Thorwart},\ and\ \citenamefont
  {Posske}}]{bonkhoff2020bosonic}%
  \BibitemOpen
  \bibfield  {author} {\bibinfo {author} {\bibfnamefont {M.}~\bibnamefont
  {Bonkhoff}}, \bibinfo {author} {\bibfnamefont {K.}~\bibnamefont
  {J\"agering}}, \bibinfo {author} {\bibfnamefont {S.}~\bibnamefont {Eggert}},
  \bibinfo {author} {\bibfnamefont {A.}~\bibnamefont {Pelster}}, \bibinfo
  {author} {\bibfnamefont {M.}~\bibnamefont {Thorwart}},\ and\ \bibinfo
  {author} {\bibfnamefont {T.}~\bibnamefont {Posske}},\ }\bibfield  {title}
  {\bibinfo {title} {Bosonic continuum theory of one-dimensional lattice
  anyons},\ }\href {https://doi.org/10.1103/PhysRevLett.126.163201} {\bibfield
  {journal} {\bibinfo  {journal} {Phys. Rev. Lett.}\ }\textbf {\bibinfo
  {volume} {126}},\ \bibinfo {pages} {163201} (\bibinfo {year}
  {2021})}\BibitemShut {NoStop}%
\bibitem [{\citenamefont {Keilmann}\ \emph {et~al.}(2011)\citenamefont
  {Keilmann}, \citenamefont {Lanzmich}, \citenamefont {McCulloch},\ and\
  \citenamefont {Roncaglia}}]{keilmann2011statistically}%
  \BibitemOpen
  \bibfield  {author} {\bibinfo {author} {\bibfnamefont {T.}~\bibnamefont
  {Keilmann}}, \bibinfo {author} {\bibfnamefont {S.}~\bibnamefont {Lanzmich}},
  \bibinfo {author} {\bibfnamefont {I.}~\bibnamefont {McCulloch}},\ and\
  \bibinfo {author} {\bibfnamefont {M.}~\bibnamefont {Roncaglia}},\ }\bibfield
  {title} {\bibinfo {title} {Statistically induced phase transitions and anyons
  in 1d optical lattices},\ }\href {https://doi.org/10.1038/ncomms1353}
  {\bibfield  {journal} {\bibinfo  {journal} {Nature communications}\ }\textbf
  {\bibinfo {volume} {2}},\ \bibinfo {pages} {1} (\bibinfo {year}
  {2011})}\BibitemShut {NoStop}%
\bibitem [{\citenamefont {Greschner}\ and\ \citenamefont
  {Santos}(2015)}]{greschner2015hubbard}%
  \BibitemOpen
  \bibfield  {author} {\bibinfo {author} {\bibfnamefont {S.}~\bibnamefont
  {Greschner}}\ and\ \bibinfo {author} {\bibfnamefont {L.}~\bibnamefont
  {Santos}},\ }\bibfield  {title} {\bibinfo {title} {Anyon hubbard model in
  one-dimensional optical lattices},\ }\href
  {https://doi.org/10.1103/PhysRevLett.115.053002} {\bibfield  {journal}
  {\bibinfo  {journal} {Phys. Rev. Lett.}\ }\textbf {\bibinfo {volume} {115}},\
  \bibinfo {pages} {053002} (\bibinfo {year} {2015})}\BibitemShut {NoStop}%
\bibitem [{\citenamefont {Hasenfratz}(1979)}]{hasenfratz1979puzzling}%
  \BibitemOpen
  \bibfield  {author} {\bibinfo {author} {\bibfnamefont {P.}~\bibnamefont
  {Hasenfratz}},\ }\bibfield  {title} {\bibinfo {title} {A puzzling
  combination: Disorder $\times$ order},\ }\href
  {https://doi.org/10.1016/0370-2693(79)91267-X} {\bibfield  {journal}
  {\bibinfo  {journal} {Physics Letters B}\ }\textbf {\bibinfo {volume} {85}},\
  \bibinfo {pages} {338} (\bibinfo {year} {1979})}\BibitemShut {NoStop}%
\bibitem [{\citenamefont {Clark}\ \emph {et~al.}(2018)\citenamefont {Clark},
  \citenamefont {Anderson}, \citenamefont {Feng}, \citenamefont {Gaj},
  \citenamefont {Levin},\ and\ \citenamefont {Chin}}]{clark2018density}%
  \BibitemOpen
  \bibfield  {author} {\bibinfo {author} {\bibfnamefont {L.~W.}\ \bibnamefont
  {Clark}}, \bibinfo {author} {\bibfnamefont {B.~M.}\ \bibnamefont {Anderson}},
  \bibinfo {author} {\bibfnamefont {L.}~\bibnamefont {Feng}}, \bibinfo {author}
  {\bibfnamefont {A.}~\bibnamefont {Gaj}}, \bibinfo {author} {\bibfnamefont
  {K.}~\bibnamefont {Levin}},\ and\ \bibinfo {author} {\bibfnamefont
  {C.}~\bibnamefont {Chin}},\ }\bibfield  {title} {\bibinfo {title}
  {Observation of density-dependent gauge fields in a bose-einstein condensate
  based on micromotion control in a shaken two-dimensional lattice},\ }\href
  {https://doi.org/10.1103/PhysRevLett.121.030402} {\bibfield  {journal}
  {\bibinfo  {journal} {Phys. Rev. Lett.}\ }\textbf {\bibinfo {volume} {121}},\
  \bibinfo {pages} {030402} (\bibinfo {year} {2018})}\BibitemShut {NoStop}%
\bibitem [{\citenamefont {G{\"o}rg}\ \emph {et~al.}(2019)\citenamefont
  {G{\"o}rg}, \citenamefont {Sandholzer}, \citenamefont {Minguzzi},
  \citenamefont {Desbuquois}, \citenamefont {Messer},\ and\ \citenamefont
  {Esslinger}}]{gorg2019realization}%
  \BibitemOpen
  \bibfield  {author} {\bibinfo {author} {\bibfnamefont {F.}~\bibnamefont
  {G{\"o}rg}}, \bibinfo {author} {\bibfnamefont {K.}~\bibnamefont
  {Sandholzer}}, \bibinfo {author} {\bibfnamefont {J.}~\bibnamefont
  {Minguzzi}}, \bibinfo {author} {\bibfnamefont {R.}~\bibnamefont
  {Desbuquois}}, \bibinfo {author} {\bibfnamefont {M.}~\bibnamefont {Messer}},\
  and\ \bibinfo {author} {\bibfnamefont {T.}~\bibnamefont {Esslinger}},\
  }\bibfield  {title} {\bibinfo {title} {Realization of density-dependent
  peierls phases to engineer quantized gauge fields coupled to ultracold
  matter},\ }\href {https://doi.org/10.1038/s41567-019-0615-4} {\bibfield
  {journal} {\bibinfo  {journal} {Nature Physics}\ }\textbf {\bibinfo {volume}
  {15}},\ \bibinfo {pages} {1161} (\bibinfo {year} {2019})}\BibitemShut
  {NoStop}%
\bibitem [{\citenamefont {Lienhard}\ \emph {et~al.}(2020)\citenamefont
  {Lienhard}, \citenamefont {Scholl}, \citenamefont {Weber}, \citenamefont
  {Barredo}, \citenamefont {de~L\'es\'eleuc}, \citenamefont {Bai},
  \citenamefont {Lang}, \citenamefont {Fleischhauer}, \citenamefont
  {B\"uchler}, \citenamefont {Lahaye},\ and\ \citenamefont
  {Browaeys}}]{lienhard2020soc}%
  \BibitemOpen
  \bibfield  {author} {\bibinfo {author} {\bibfnamefont {V.}~\bibnamefont
  {Lienhard}}, \bibinfo {author} {\bibfnamefont {P.}~\bibnamefont {Scholl}},
  \bibinfo {author} {\bibfnamefont {S.}~\bibnamefont {Weber}}, \bibinfo
  {author} {\bibfnamefont {D.}~\bibnamefont {Barredo}}, \bibinfo {author}
  {\bibfnamefont {S.}~\bibnamefont {de~L\'es\'eleuc}}, \bibinfo {author}
  {\bibfnamefont {R.}~\bibnamefont {Bai}}, \bibinfo {author} {\bibfnamefont
  {N.}~\bibnamefont {Lang}}, \bibinfo {author} {\bibfnamefont {M.}~\bibnamefont
  {Fleischhauer}}, \bibinfo {author} {\bibfnamefont {H.~P.}\ \bibnamefont
  {B\"uchler}}, \bibinfo {author} {\bibfnamefont {T.}~\bibnamefont {Lahaye}},\
  and\ \bibinfo {author} {\bibfnamefont {A.}~\bibnamefont {Browaeys}},\
  }\bibfield  {title} {\bibinfo {title} {Realization of a density-dependent
  peierls phase in a synthetic, spin-orbit coupled rydberg system},\ }\href
  {https://doi.org/10.1103/PhysRevX.10.021031} {\bibfield  {journal} {\bibinfo
  {journal} {Phys. Rev. X}\ }\textbf {\bibinfo {volume} {10}},\ \bibinfo
  {pages} {021031} (\bibinfo {year} {2020})}\BibitemShut {NoStop}%
\bibitem [{\citenamefont {Yao}\ \emph {et~al.}(2022)\citenamefont {Yao},
  \citenamefont {Zhang},\ and\ \citenamefont {Chin}}]{yao2022domain}%
  \BibitemOpen
  \bibfield  {author} {\bibinfo {author} {\bibfnamefont {K.-X.}\ \bibnamefont
  {Yao}}, \bibinfo {author} {\bibfnamefont {Z.}~\bibnamefont {Zhang}},\ and\
  \bibinfo {author} {\bibfnamefont {C.}~\bibnamefont {Chin}},\ }\bibfield
  {title} {\bibinfo {title} {Domain-wall dynamics in bose--einstein condensates
  with synthetic gauge fields},\ }\href
  {https://doi.org/10.1038/s41586-021-04250-3} {\bibfield  {journal} {\bibinfo
  {journal} {Nature}\ }\textbf {\bibinfo {volume} {602}},\ \bibinfo {pages}
  {68} (\bibinfo {year} {2022})}\BibitemShut {NoStop}%
\end{thebibliography}%

\end{document}